\documentclass[twocolumn,showpacs,showkeys,amsmath,amssymb]{revtex4}
\usepackage{graphicx}% Include figure files
\usepackage{dcolumn}% Align table columns on decimal point
\usepackage{bm}% bold math\begin{document}

\usepackage{mathtext}
\usepackage{graphics}
\usepackage[dvips]{epsfig}
\usepackage{epsfig}
\usepackage{subfigure}
\usepackage{mathrsfs}
\usepackage{amssymb,latexsym}
\def\shiftleft#1{#1\llap{#1\hskip 0.04em}}
\def\shiftdown#1{#1\llap{\lower.04ex\hbox{#1}}}
\def\thick#1{\shiftdown{\shiftleft{#1}}}
\def\b#1{\thick{\hbox{$#1$}}}
\begin{document}
%\preprint{Moscow-T\"ubingen 2006}

\title{Isoscalar short-range current in the deuteron\\
induced by an intermediate dibaryon}

\thanks{This work was partially supported by the Russian 
Foundation for Basic Research (grants 05-02-04000 and 05-02-17407) 
and the Deutsche Forschungsgemeinschaft (grant Fa-67/20-1).}

\author{V.I. Kukulin, I.T. Obukhovsky,}%
\affiliation{Institute of Nuclear Physics, Moscow
State University,119899 Moscow, Russia.}

\author{P. Grabmayr,}%
\affiliation{Physikalisches Institut, Universit\"at T\"ubingen, 
D72076 T\"ubingen, Germany}%

\author{and Amand Faessler}%
\affiliation{Institut f\"ur Theoretische Physik, 
Universit\"at T\"ubingen, D72076 T\"ubingen, Germany}%

\date{\today}%

\begin{abstract}
  A new model for short-range isoscalar currents in the deuteron and in the
  $NN$ system is developed; it is based on the generation of an intermediate
  dibaryon which is the basic ingredient for the medium- and short-range $NN$
  interaction which was proposed recently by the present authors. The new
  current model has only one parameter, which moreover has a clear physical
  meaning.  Our calculations have demonstrated that this new current model can
  very well describe the experimental data for the three basic deuteron
  observables of isoscalar magnetic type, viz. the magnetic moment, the
  circular polarization of the photon in the $\vec{n}p\to d\vec\gamma$ 
  process at thermal neutron energies and the structure function $B(Q^2)$ 
  up to $Q^2\simeq$ 60~fm$^{-2}$.
\end{abstract}

\maketitle

%%%%%%%%%%%%%%%%%%%%%%%%%%%%%%%%%%%%%%%%%%%%%%%%%%%%%%%%%%%%%%%%%%%%%%
\section{\label{sec:level1} Introduction.}
%%%%%%%%%%%%%%%%%%%%%%%%%%%%%%%%%%%%%%%%%%%%%%%%%%%%%%%%%%%%%%%%%%%%%%%%

The problem of electromagnetic currents in deuteron, especially of
isoscalar nature, still cannot be considered as fully resolved 
despite of very numerous and longstanding efforts of many groups.
As a good example we name here three topics from the field where the
present approaches all relying on the conventional nucleon-nucleon
($NN$) force models, failed to explain quantitatively the experimental 
deuteron data. These are:
\begin{itemize}
\item[($i$)] the circular polarization $P_{\gamma}$ of $\gamma$-quanta 
in radiative capture of spin-polarized neutrons in hydrogen at thermal 
energies~\cite{lob,mul}, viz. $\vec n+p\to d+\vec\gamma$ reaction;
\item[($ii$)] the deuteron magnetic form factor $B(Q^2)$ around the 
diffraction minimum at $Q_{min}\sim$ 50 
$fm^{-2}$~\cite{sac,slac,jlab,phil,tjon};
\item[($iii$)]the photon-induced polarization of the neutron 
from the $d(\gamma,\vec n)p$ reaction at low energy\cite{sci}.
\end{itemize}

When going to few-nucleon electromagnetic observables the disagreements
of modern models for two-body currents with existing experimental data becomes even
far more numerous~\cite{blun}. The main part of the above discrepancies
is related to the isoscalar magnetic currents which are now still
strongly model dependent~\cite{phil,tjon,sci,blun,arenh,carl}, and thus 
cannot be established uniquely. 
Moreover, the existing theoretical approaches seems to employ
all the important types of currents, viz. one-nucleon, two- and 
three-nucleon ones with many types of meson-exchange currents 
($\rho\pi\gamma$, $\omega\pi\gamma$, etc.) to remove the  
discrepancies. So, the various discrepancies 
which still remained could imply likely that some important
contributions to the electromagnetic currents are still absent. These
current components ignored until now are able to remove at least some
of the disagreements observed to date in the deuteron- and few-nucleon
electromagnetic observables.

In the present paper we propose a model for new currents of 
isoscalar nature, which still have not been discussed up to date. 
It is demonstrated below that the new current removes the 
discrepancies enumerated above in ($i$) - ($ii$) that they make the
theoretical framework for electromagnetic properties of the deuteron
and the few-nucleon systems more consistent and thus more powerful. 

Hence, we shall discuss in Sect.~\ref{sec:currents} briefly the status of 
isoscalar magnetic currents in the deuteron and in the $NN$ system at low 
energies (see more extended reviews 
in~\cite{arenh,carl,ris,marcu,rho4,che}).
The dibaryon model for the $NN$ interaction is elaborated in
Sect.~\ref{sec:dressedbag} where we also include both EFT- and microscopic
quark shell-model approaches to substantiate our model. In
Sect.~\ref{sec:deuteron} the properties of deuteron as emerging from the new
force model are discussed in detail. We describe the short-range currents
induced by the intermediate dibaryon component and the microscopic quark model
formalism for the deuteron electromagnetic currents.
Sect.~\ref{sec:transampl} is devoted to a general consideration of the
isoscalar $M1$- and $E2$-transition amplitudes and we present in detail the
formalism for the isoscalar currents. In particular the formalism for the
calculation of the $\vec{n}p{\to}d\gamma$ reaction, the form factor $B(Q^2)$
and the magnetic moment $\mu_d$ is discussed.  The numerical results and their
comparison to the respective experimental data are presented in
Sect.~\ref{sec:experiment}.  The summary of the results obtained is given in
Sect.~\ref{sec:discussion}.  In the Appendices some useful formulas and some
details of the derivation have been collected.

%%%%%%%%%%%%%%%%%%%%%%%%%%%%%%%%%%%%%%%%%%%%%%%%%%%%%%%%%%%%%%%%%%%%%%%%%%%%%%%
\section{The status of isoscalar currents in the deuteron}
\label{sec:currents}
%%%%%%%%%%%%%%%%%%%%%%%%%%%%%%%%%%%%%%%%%%%%%%%%%%%%%%%%%%%%%%%%%%%%%%%%%%%%%%%

A consistent and correct description of isoscalar currents in the 
deuteron and in the $NN$-system at very low energy 
seems to open a door toward quantitative studies of the
short-range $NN$-force at low energies. Through this also toward
to the non-perturbative QCD  at low energies where there are no 
extra difficulties encountered compared to investigations at high energies,
such as a necessity for relativistic treatment, incorporation of many 
inelastic processes etc. The problem has been first posed 
by Breit and Rustgi~\cite{breit} many years ago. Main part of the modern
efforts to understand the problem quantitatively comes from effective
field theory (EFT) treatment in high-orders (N$^3$LO, N$^4$LO etc.)
and also from the model treatment of short-range $NN$-interaction
within phenomenological $NN$-potentials. The point is that the main
contribution to the amplitude of processes like $n\!+\!p\to d\!+\!\gamma$
comes from the isovector M1 current. This two-body current is dominated
by a long-range $\pi$-meson exchange. This phenomenon has been named
in literature as a chiral filter hypothesis~\cite{rho1}. So, the
isovector M1-transitions are ``protected by the chiral filter'' and
does not manifest any sensitivity to short-range interactions. 

In the sharp contrast to this, two-body {\it isoscalar} current  is not 
protected by the chiral filter and thus depends sensitively on the short-range 
interactions and short-range current contributions. This renders such
currents to be highly model dependent. As a result, within the framework 
of EFT-approach the treatment of isoscalar currents requires some
high orders of ChPT-expansion (N$^3$LO and N$^4$LO), which, in turn,
introduces some extra parameters~\cite{rho4,che,rho2,rho3}. On the
other hand, in a more conventional MEC-treatment (see e.g. 
\cite{blun,carl,ris}) the quantitative description of isoscalar 
currents includes, in addition to the usual $\rho\pi\gamma$-contribution, 
also the relativistic corrections dependent on the $NN$
interaction model. All these contributions depend
crucially upon the meson-nucleon form factor cutoffs $\Lambda_{\pi NN}$,
$\Lambda_{\pi N\Delta}$, $\Lambda_{\rho NN}$, $\Lambda_{\omega NN}$ etc.
and also upon the electromagnetic form factors of the intermediate 
mesons\cite{arenh}.
It is important to emphasize that the above cutoff values chosen for
the two-body current models are generally not the same as in the input
of $NN$ and 3$N$ potential models~\cite{sasa} and rather often are simply 
fitted to the electromagnetic observables measured in experiments. 
The clear evidence to the strongly enhanced values for the cutoffs
$\Lambda_{\pi NN}$, $\Lambda_{\pi N\Delta}$, $\Lambda_{\omega NN}$ etc. 
chosen in traditional $NN$-models like the Bonn potential model can be 
seen in the results of relativistic calculations for the deuteron structure
functions $A(q^2)$ and $B(q^2)$, and the deuteron magnetic 
moment~\cite{arenh}. 
%In fact, an addition of the relativistic corrections
%to the non-relativistic results (using the potential model with very large
%cutoff values) leads to strong overshooting for all the deuteron form 
%factors at $q^2>$ 20 fm$^{-2}$ and also for the magnetic moment. 
It is very likely that a serious overestimation of the relativistic
contributions found in Ref.~\cite{arenh} indicates to the too high
momentum cutoffs in the meson-nucleon vertices.

Thus, one can summarize that the quantitative treatment of isoscalar 
currents in all existing approaches is related to the rather strong model 
dependence, in contrast to the isovector M1 current.
For example, the experimental value for the circular 
polarization $P_{\gamma}$ of $\gamma$-quanta emitted in the 
$np\to d\gamma$ process measured some times ago~\cite{lob} is
$P_{\gamma}^{\,exp}\simeq (-1.5\pm0.3)\times10^{-3}$
and is underestimated by all existing $NN$- and two-body current
models~\cite{khrip}.
However, despite of some inner inconsistencies in the 
two-body current models mentioned above it occurred to be feasible to 
describe many deuteron electromagnetic 
observables like static characteristics, charge and 
quadrupole form factors, the $np\to d\gamma$ cross sections at low 
energies etc., rather accurately. 
But the description of other electromagnetic observables, in particularly
those given above in the items ($i$) - ($iii$), meets quite serious
difficulties. 

In particular, there is a long standing puzzle tightly interrelated 
to the deuteron isoscalar current 
is a behavior of the deuteron magnetic form factor 
$B(Q^2)$ in the area around the diffraction minimum, $Q^2\sim$ 45 - 55
$fm^{-2}$. There is a huge literature devoted to calculations 
of the deuteron form factors. However, the most recent
fully relativistic treatment of the $B(Q^2)$-behavior~\cite{phil} has clearly
demonstrated that the existing two-body current models have missed some
important contributions. Missing the same contributions very likely is the
reason for the visible disagreement of vector and tensor analyzing powers in
$p$-$d$ and $n$-$d$ radiative capture at very low energies.  Thus, it is
evident, there is a number of mutually interconnected effects in the deuteron
and the few-nucleon systems where one needs a new isoscalar current. However,
the conventional two-body current models still have no appropriate candidates
for this.

In this paper we propose an alternative mechanism for isoscalar 
current generation in the two-nucleon system based on the dibaryon model
for $NN$ interactions at intermediate and short ranges developed recently
by our group~\cite{kuk1,kuk2,kuk6,kuk7}. Generally speaking, employment 
of the dibaryon degree of freedom to describe quantitatively the  
electromagnetic
deuteron properties at low energies is not new. As an example, we shall 
refer to numerous studies having appeared in 80-ies where authors tried to 
incorporate various types of six-quark bags to the deuteron wave
function to explain some puzzles observed in electromagnetic observables 
(see, e.g. Refs.~\cite{yama,oka,kissl,buch,kobu}).
These early attempts have revealed the important role which the quark 
degrees of freedom plays in the short-range $NN$ interaction
and in the electromagnetic structure of deuteron (see, e.g., \cite{kobu}
and references therein). The very important role played by a
``hidden color'' and the quark pair currents in the deuteron electromagnetic
form factors has been established in these works, in particular.  However,
these early attempts have resulted neither in a deeper understanding of the
deuteron electromagnetic observables nor in significant improvement of the
description of experimental data.

Very recently, however, in tight interrelation to a renaissance of 
the dibaryon physics (which is related partially to a recent boom in
pentaquark physics), many new studies, both 
theoretical~\cite{wang,pang,goldm,pang1,ardo,beane,flem}
and experimental~\cite{khry,khry1,zoln}, 
have appeared in literature. An EFT-approach which included intermediate 
dibaryons~\cite{ardo,beane,flem} focused
just on description of electromagnetic properties of the deuteron at 
low energies
and momentum transfer. However, the dibaryons in that approach have
been introduced more on formal basis than as a real physical degree 
of freedom (or an intermediate broad resonance), to improve the
convergence of chiral perturbation series and to simplify the EFT
scheme in that case. In our dibaryon approach, in contrast to the formal 
scheme~\cite{wang,pang,goldm,pang1}, the dibaryons are considered 
as an important new degree of freedom in the description of 
$NN$- and 3$N$ interactions in nuclei at 
intermediate and short ranges~\cite{kuk1,kuk2,kuk7}.

The dibaryon concept of the nuclear force at intermediate and short ranges
turned out very fruitful in a quantitative or qualitative description
of numerous phenomena in hadron and nuclear physics. In particular,
this model made it possible to explain without any free parameters the 
Coulomb displacement energy for $A=$ 3 nuclei ($^3H$ and $^3He$) and
all ground-state characteristics of these nuclei~\cite{kuk5}; 
with exception of the $^3$H and
$^3$He magnetic moments where we included only single-nucleon currents.
However, according to general principle of quantum mechanics, any new degree
of freedom must lead to new currents of diagonal and transitional type. On the
other hand, in our approach the weight of the dibaryon component in deuteron
does not exceed 2.5 - 3.5\%. Therefore, just the transitional current from the
$NN$ to the dibaryon component can be more important than the diagonal
dibaryon current. Nevertheless, we have considered in the present study the
diagonal dibaryon current as well.

%%%%%%%%%%%%%%%%%%%%%%%%%%%%%%%%%%%%%%%%%%%%%%%%%%%%%%%%%%%%%%%%%%%%%     
\section{The dressed bag model for the medium- and 
short-range  $NN$ interaction}
\label{sec:dressedbag}
%%%%%%%%%%%%%%%%%%%%%%%%%%%%%%%%%%%%%%%%%%%%%%%%%%%%%%%%%%%%%%%%%%%%%%

A few key points have been taken as the physical justification of this 
dibaryon model.
\begin{itemize}
\item[($i$)] Failure of the $\sigma$-meson exchange to describe the 
strong intermediate-range 
attraction in the $NN$ channel~\cite{kais,oset,kasku}. 
Instead of a strong 
attraction required by the $NN$-phase shifts, the exchange of two 
correlated pions with the $S$-wave broad $\sigma$-meson resonance
in $\pi$-$\pi$ scattering included to the mechanism, when treated 
consistently and with a rigor, leads to a strong
medium-range {\em repulsion} and a weak long-range attraction.
Thus, the crucial problem arises, how to explain the basic strong 
attraction between two nucleons at intermediate range.
\item[($ii$)] The heavy-meson exchange with the mass $m_{\rho,\omega}\sim$
800 MeV corresponds in general to the inter-nucleon distances 
$r_{NN}\sim$ 0.3 fm, i.e. it occurs deeply inside the overlap region 
between two nucleons. And thus, these exchanges should be consistently
treated only with invoking a six-quark dynamics. The ignorance of
this dynamics in conventional $NN$-potential approaches (of the OBE 
type) leads to several serious inconsistencies in
OBE-description of short-range $NN$ force (see the discussion in 
Refs.~\cite{kuk1,kuk2,krebl}) and is also a reason for some problems with 
description of the short-range $NN$ correlation in nuclei.
\item[($iii$)] One more serious difficulty is related to the choice 
of different cutoff parameter values $\Lambda_{\pi NN}$, 
$\Lambda_{\pi N\Delta}$ etc., when one set is used
to describe the elastic $NN$ scattering, another set 
to fit the pion production cross section in $NN$ scattering and
the meson exchange contribution to the two-body current,
and a third set to describe the three-body 
force~\cite{carl,kuk1,kuk2}. Using these different values for the identical
vertices in the description of tightly interrelated quantities
points toward an inadequate framework employed in the underlying 
model.
\end{itemize}
The dibaryon model~\cite{kuk2,kuk7} seemingly overcomes  the above 
difficulties with a consistent description of the short-range $NN$ force.
Firstly, it employs a rather soft cutoff parameter $\Lambda_{\pi NN}$
(moreover, there is likely no serious sensitivity to these cutoffs in
our model). Secondly, the short-range $NN$ interaction is described
through a dressed dibaryon dynamics, so that the conventional 
heavy-meson exchange 
in $t$-channel plays no serious role in the short-range $NN$-dynamics.
And thirdly, the short-range $NN$-correlations in nuclei
tested by high-energy probes can consistently be described as an 
interaction of the probes with the dibaryon as a whole, which may be
associated with its inner excitations or de-excitations. 

\begin{figure}[hp]\centering
\epsfig{file=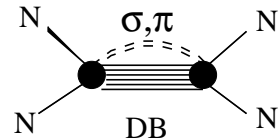,width=0.2\textwidth}
\caption{Driving mechanism for the dressed dibaryon generation used as
an intermediate state in the short-range $NN$-interaction.}
\label{f3}
\end{figure}

In our approach the virtual dibaryon in the $NN$ system are modeled
through generation of a symmetric six-quark configuration 
$s^6[6]_{\scriptscriptstyle X}$ dressed with a strengthened scalar 
(e.g. $\sigma$-meson) and other (e.g., $\pi$, $\rho$, $\omega$ etc.) fields. 
It can be schematically illustrated by Fig.~\ref{f3}.
%%%%%%%%%%%%%%%%%%%%%%% Fig. 1 %%%%%%%%%%%%%%%%%%%%%%%%%%%%%%%%%%%%%%
%------------------------------------------------------------------------------
%------------------------------------------------------------------------------
%%%%%%%%%%%%%%%%%%%%%%%%%%%%%%%%%%%%%%%%%%%%%%%%%%%%%%%%%%%%%%%%%%%%%%
The enhancement of the scalar field in the
symmetric six-quark configuration $s^6[6]_{\scriptscriptstyle X}$  
implies some rearrangement of quark-gluon fields in the region, 
where two nucleons are totally overlapping. The emergence of a strong 
scalar-isoscalar field in the six-quark bag induces 
automatically an isoscalar exchange current in the multi-quark system.

The $\sigma$-meson loops originate mainly from  ``non-diagonal''
transitions from the
mixed-symmetry 2$\hbar\omega$-excited configurations
$|s^4p^2[f_X][f_{CS}]\rangle$ to the unexcited fully symmetric configuration
$|s^6[6_X]\rangle$ in the six-quark system with emission of a 
(virtual or real) $\sigma$-meson.
In turn, the strongly attractive interaction between the 
scalar-isoscalar $\sigma$-field and the multi-quark bag results in
an enhancement of the attractive very short-range diquark
correlations in the multi-quark system. Thus, as a net result of
all these highly non-linear effects the mass of the intermediate 
dibaryon surrounded by the $\sigma$-field gets much lower as
compared to the respective bare dibaryon (see the discussion below).
 
Completely similar to the behavior of nucleonic system in a scalar field, 
where the scalar exchange may be viewed as a renormalization of the nucleon
mass~\cite{Serot} 
\[m\to m^*=m+v_s(0)\rho,\]
where $v_s(0)$ is a volume integral of the scalar exchange potential and 
$\rho$ is the nucleon density, the constituent quark mass in multi-quark 
system is also reduced due to interaction with the scalar field:
\[v_s(0)\sim-g^2_{\sigma q}/m_{\sigma}^2,\]
where $g_{\sigma q}$ is the $\sigma$-quark coupling constant; note
$v_s(0)<$ 0 . Thus, the 
dibaryon
mass should be renormalized noticeably due to this scalar field mechanism.

The effect of the strong attraction of the $\sigma$-field to the quark core 
and the resulting mass shift of the multi-quark system is illustrated easily
by the anomalously low mass of the Roper 
resonance $N^*$(1440) with positive parity, which lies even lower
than the lowest negative parity resonance $N^*$(1535). 
It is widely accepted~\cite{krebl} that the Roper resonance has mainly
the structure $|N+\sigma\rangle$. However, in the language of the simple 
quark shell model, the second positive parity level in the nucleon 
spectrum corresponds to 2$\hbar\omega$ excited three-quark 
configurations $|sp^2[21]_X\rangle$ or $|s^2d-(s^22s)[21]_X\rangle$. Thus, 
exactly the same 
mechanism as in our dibaryon model, i.e. the $\sigma$-meson emission
from the 2$\hbar\omega$-excited state of the nucleon 
$|sp^2[21]_X\rangle\to |s^3[3]_X+\sigma\rangle$, leads to a generation of 
a strong $\sigma$-field
and to a significant shift downward of the Roper resonance 
mass. Without this strong non-linear effect, the mass of the
$N^*$(1440) would be $\sim$ 500 MeV higher than that found
experimentally. This very large shift gives some estimate
for the magnitude of the attractive effects which appear in the interaction 
of the scalar field with fully symmetric multi-quark bag, or in
other words, the effect of dressing by the $\sigma$-meson field. 
At very short $NN$ distances the $qq$ correlations become repulsive
due to one-gluon exchange, and jointly with a strongly 
enhanced quark kinetic energy, this results in effective short-range
repulsion in the $NN$-channel. In our approach this repulsive 
part of non-local $NN$ interaction is modeled by a separable term
$\lambda_0\varphi_{0S}({\bf r})\varphi_{0S}({\bf r}^{\prime})$ 
with a large positive coupling constant $\lambda_0$~\cite{kuk1,kuk2}
while the form factor $\varphi_{0S}({\bf r})$ is the projection of the 
six-quark $|s^6[6]_X\rangle$ state onto the $NN$-channel: 
$\varphi_{0S}({\bf r})=\langle N\,N|s^6[6]_X\rangle$.

Such combined mechanism lies 
fully beyond the perturbative QCD, and we suggest 
it can be described phenomenologically by
dressing the six-quark propagator $G_{6q}(E)$ with 
$\sigma$-meson loops~\cite{kuk6,kuk7}. The resulting dressed-bag 
(DB) propagator $G_{DB}(E)$ and the transition vortexes $NN\to DB$ 
and $DB\to NN$ treated  within a microscopic $6q$ model~\cite{kuk2}
lead automatically to a non-local 
(separable) energy-dependent short-range $NN$ potential 
$V_{NqN}(r,r^{\prime};E)$
\begin{eqnarray}
V_{NqN}&\equiv&\langle NN|V_{Nq}|DB(s^6)\rangle G_{DB}(E)\,
\langle DB(s^6)|V_{qN}|NN\rangle\nonumber\\
&=&\varphi(r)\lambda(E)\varphi(r^{\prime}),
\label{sprb}
\end{eqnarray}
where the form factors $\varphi(r)$ are deduced from the microscopic
$6q$ model while the coupling constant $\lambda(E)$ is determined by
a loop integral with the $\sigma$-loop as shown schematically in 
Fig.~\ref{f3} \cite{kuk2}. This loop integral can be conventionally
parametrized in the following Pade form
\begin{eqnarray}
\lambda(E)=\lambda(0)\frac{E_0+aE}{E_0-E}.
\label{lame}
\end{eqnarray}
Here the parameters $\lambda(0)$, $E_0$ and $a$
can be either calculated from the microscopic $6q$ model or obtained
from fits
to the phase shifts of $NN$ scattering in the low partial waves.
This approach resulted in the Moscow-Tuebingen (MT) 
potential model of $NN$ interaction~\cite{kuk1,kuk2}.

\begin{figure}[hp]\centering
\epsfig{file=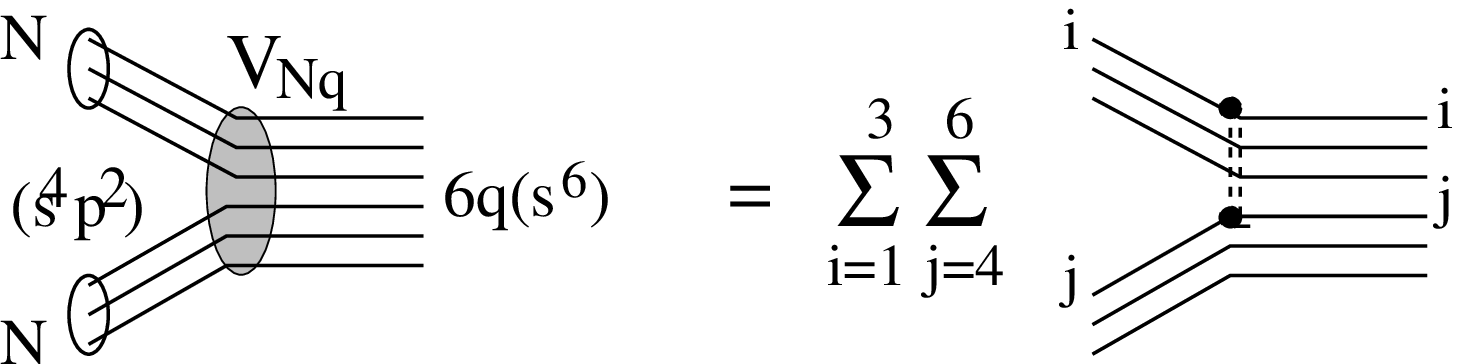,width=0.4\textwidth}
\caption{Graphical illustration for $NN\to DB$ transition in terms 
of quark microscopic model. The double-dashed line denotes some scalar
exchange which can induce the $NN\to DB$ transition~\cite{kuk6,kuk7}.}
\label{f4}
\end{figure}

In a more general treatment recently developed in Refs.~\cite{kuk6,kuk7}
the one-pole approximation for $V_{NqN}$ was obtained on the basis of
a fully covariant EFT approach.
In a more simple version of the model~\cite{kuk1,kuk2}
the transition operator $V_{Nq}$ which couples the nucleon-nucleon
and dibaryon channels was calculated within a microscopic quark
model with employment of the quark-cluster decomposition of the 
short-range $N$-$N$ wave function. We consider here the $NN\to DB$ 
transition in terms of this simple model, shown schematically 
in Fig.~\ref{f4}, where we assume that the coupling between the $NN$ 
and $DB$ channels is realised on the quark level
via a scalar exchange. This scalar interaction can be presented not only 
through the $\sigma$-meson exchange but also through a quark confinement 
or another force including even the four-quark instanton-induced interaction 
of t'Hooft's type~\cite{crist}. 

%%%%%%%%%%%%%%%%%%%%%%% Fig. 2 %%%%%%%%%%%%%%%%%%%%%%%%%%%%%%%%%%%%%%
%------------------------------------------------------------------------------
%------------------------------------------------------------------------------
%%%%%%%%%%%%%%%%%%%%%%%%%%%%%%%%%%%%%%%%%%%%%%%%%%%%%%%%%%%%%%%%%%%%

The transition operator $V_{Nq}$ can be written in the form
\begin{eqnarray}
V_{Nq}=\sum_{i\!=\!1}^3\sum_{j\!=\!4}^6g_s^2v_s(r_{ij}),
\label{scl}
\end{eqnarray}
where $v_s(r_{ij})$ is a scalar $qq$ interaction. This operator should
be substituted into the transition matrix element in Eq.~(\ref{sprb}). 
The particular form of the scalar operator~(\ref{scl}) and its
origin are not significant here. The specific mechanism of dressing
can also be disregarded at this step since the 
dressing has already been taken into account in the propagator
$G_{DB}$. The most important point is that 
the S-wave two-cluster state $3q+3q$, which represents the $N$-$N$
system in the overlap region, should be very close in its symmetry
to a superposition of the excited six-quark configurations 
$s^4p^2$ (see Refs.~\cite{harv,obu1} for detail) having non-trivial 
permutation 
symmetries with the following Young tableaux in the coordinate- and 
color-spin spaces
$\{f\}\equiv\{[f_{\scriptscriptstyle X}],[f_{\scriptscriptstyle CS}]\}$
($[f_{\scriptscriptstyle X}]=[6],\,[42], [f_{\scriptscriptstyle CS}]
=[42],[321],[2^3],[31^3],[21^4]$). 
For this reason, we can rewrite the effective $NN$ interaction 
Eq.~(\ref{sprb})
generated by the transition operator Eq.~(\ref{scl}) in the following 
constrained form:
\begin{widetext}
\begin{multline}
V_{NqN}(r,r^{\prime};E)\simeq\sum_{ff^{\prime}}\{\langle N(123)|\langle 
N(456)|\}|s^4p^2\{f\}\rangle\,\langle s^4p^2\{f\}|V_{Nq}
|s^6[6]_{\scriptscriptstyle X}\rangle\\
\times G_{DB}(E)\,\langle s^6[6]_{\scriptscriptstyle X}|V_{Nq}|
s^4p^2\{f^{\prime}\}\rangle\,\langle s^4p^2\{f^{\prime}\}|
\,\{|N(123)\rangle |N(456)\rangle \},
\label{cnstr}
\end{multline}
\end{widetext}
where we leave the mixed-symmetry 2$\hbar\omega$-excited six-quark 
configuration $s^4p^2$ only (but with all the possible Young 
tableaux $\{f\}$), instead of the total sum over all the 
excited six-quark configurations $s^4p^2$, $s^2p^4$, \dots, etc.
Then, one can deduce from Eq.~(\ref{cnstr}) that the matrix element of 
the $NN\to DB$ transition is proportional to the wavefunction of an 
{\it excited} nucleon-nucleon $2S$-state: 
\begin{eqnarray}
\{\langle N(123)|\langle N(456)|\}\,|\,s^4p^2\{f\}\rangle \,
=C_f\,\varphi_{2S}(r).
\label{prj}
\end{eqnarray}
Here $C_f$'s are purely algebraic coefficients and 
${\bf r}=({\bf r}_1+{\bf r}_2+{\bf r}_3-{\bf r}_4-{\bf r}_5-{\bf r}_6)/3$ 
is the relative-motion coordinate of the two $3q$-clusters. 
For simplicity, we use here 
the harmonic oscillator (h.o.) $2S$-state function
$\varphi_{2S}(r)=2r_0^{-3/2}\pi^{-1/4}\sqrt{3/2}
\left(2r^2/3r_0^2-1\right)exp(-r^2/2r_0^2)$ for the projection
of the mixed-symmetry six-quark state onto the $NN$-channel, 
which is characteristic of the quark shell model. The nucleon wave
function N(ijk) in such an approach is described by the pure $s^3$ 
configuration of the CQM 
\begin{multline}
|N(123)\rangle\,=|s^3[3]_{\scriptscriptstyle X}
\{f_{\scriptscriptstyle ST}\}\rangle\\
=\psi_N(\rho_1,\rho_2)|
[2^3]_{\scriptscriptstyle C}[3]_{\scriptscriptstyle ST}\rangle,
\label{n123}
\end{multline}
where $\psi_N(\rho_1,\rho_2)={\cal N}exp\left[
-\frac{1}{2b^2}(\rho_1^2/2+2\rho_2^2/3)\right]$, the parameter
$b$ is the scale parameter of the CQM, with $b\simeq$ 0.5 - 0.6 fm;
the relative coordinates are 
${\b\rho}_1={\bf r}_1-{\bf r}_2$ and 
${\b\rho}_2=({\bf r}_1+{\bf r}_2)/2-{\bf r}_3$.

Then, using the 2S function for the transition $NN\to DB$ vertex
in Eq.~(\ref{sprb}), i.e. substituting $\varphi(r)=\varphi_{2S}(r)$, 
we obtain in our simple ansatz Eq.~(\ref{scl}) for the 
$qq$ pair interaction $v_s(r_{ij})$
\begin{multline}
\langle NN(s^4p^2)|V_{Nq}|DB(s^6)\rangle\,=g_{s}^2\,
\langle v\rangle\,\varphi_{2S}(r),\\
\mbox{and}\quad \lambda(E)=g_{s}^4\,\langle v\rangle ^2G_{DB}(E),
\label{2s}
\end{multline}
where $\langle v\rangle $ is a superposition (with the algebraic 
coefficients $C_f$) 
of the quark shell-model {\it transition matrix elements} 
$\,\langle s^4p^2|\sum_{i\!=\!1}^3\sum_{j\!=\!4}^6v_s(ij)|s^6\!\rangle$.

Note that we did not include the projector
$|s^6[6]_{\scriptscriptstyle X}\rangle \langle s^6[6]_{\scriptscriptstyle X}|$
onto the lowest six-quark configuration $s^6[6]_X$
in the sum within Eq.~(\ref{cnstr}),
because this configuration, according to the most recent $6q$ 
calculations~\cite{stan} lies well above 2.5 GeV and it is not
touched by a strong renormalization, as does the mixed-symmetry state
due to strong coupling to the $\sigma$-meson field. Thus, 
the strong repulsive contribution from the bare intermediate 
$s^6$ bag to the effective short-range $NN$ interaction
in the Moscow-Tuebingen (MT) model~\cite{kuk1,kuk2} is modeled by
an orthogonality condition to the nodeless 0S state:
\begin{eqnarray}
\int\psi_{NN}(r)\varphi_{0S}(r)d^3r=0, 
\label{ort}
\end{eqnarray}
where $\varphi_{0S}$ is a projection of the $6q$-bag state onto 
the product of nucleon wavefunctions: 
\begin{eqnarray}
\varphi_{0S}(r)={\cal N}_0^{-1}<\!N(123)|
\langle \!N(456)|\}|s^6[6]_{\scriptscriptstyle X}\rangle .
\label{0spr}
\end{eqnarray}
Through the orthogonality constraint (\ref{ort}) - (\ref{0spr}) the 
symmetric six-quark bag configuration 
is excluded from the $NN$ Hilbert space, which prevents a possible double
counting of the $s^6$ configuration. As a result, the total wavefunction 
of the two-nucleon system $\Psi_{tot}$ is defined in the
direct sum of two Hilbert spaces ${\cal H}_{NN}\oplus{\cal H}_{DB}$.
This direct sum can be conventionally 
represented by the two-line Fock column. For example, the 
deuteron state in the MT-model reads
\begin{eqnarray}
|d\rangle\, =\left(\begin{array}{c}cos\,\theta_{Nq}|d(NN)\rangle \\
\sin\theta_{Nq}|DB\rangle \end{array}\right),
\label{fock}
\end{eqnarray}
where the mixing angle $\theta_{Nq}$ is calculated on the basis of
coupled channel equations with the transition operator
$V_{Nq}$ taken as a coupling potential~\cite{kuk5}. Here the deuteron 
wavefunction in the $NN$-channel (or the $NN$-component of the 
deuteron) takes the conventional form
\begin{eqnarray}
|d(NN)\rangle\, =\frac{u(r)}{r}{\cal Y}^{01}_{1M}(\hat r)
+\frac{w(r)}{r}{\cal Y}^{21}_{1M}(\hat r),
\label{dnn}
\end{eqnarray}
where the S-wave component $u(r)$ satisfies to the constraint:
\begin{eqnarray}
\int_0^{\infty}u(r)\varphi_{0S}(r)dr=0,
\label{ortcnstr}
\end{eqnarray}
The deuteron $NN$-component $|d(NN)\rangle $ and the dibaryon component
$|DB\rangle$ are normalized individually to 1.  This implies a standard
normalization of the total wavefunction
\begin{multline}
\langle d|d\rangle\, =\\
cos^2\theta_{Nq}\langle d(NN)|d(NN)\rangle 
+sin^2\theta_{Nq}\langle DB|DB\rangle\ = 1 
\end{multline}
\noindent

The dressed dibaryon propagator $G_{DB}(E)$ can be represented through 
the Dyson equation:
\begin{eqnarray}
G_{DB}=G^{(0)}_{DB}+G^{(0)}_{DB}\Sigma\, G_{DB},
\label{dyson}
\end{eqnarray}
where $G^{(0)}_{DB}$ is the bare dibaryon propagator and the $\Sigma$
is an eigenenergy which includes irreducible diagrams, 
%%(see Figs.~\ref{fdy})
 i.e. those which do not include the intermediate
free nucleon lines (but still can include $N\Delta$ or $\Delta\Delta$
channels). Our $DB$-model calculations kept
only the leading $\sigma$-loop~\cite{kuk6,kuk7}
in the series for the $\Sigma$-kernel, because all other graphs 
(calculated within the
six-quark shell model for the quark bag) corresponds to much higher
masses (for the details see Refs.~\cite{kuk6,kuk7}). 
Moreover, the propagator 
for the bare six-quark bag $G^{(0)}_{DB}$ corresponds to the pure 
six-quark bag states with 0$\hbar\omega$ and 2$\hbar\omega$ quanta
for even partial waves and 1$\hbar\omega$ for odd partial waves.

When including the $\sigma$-meson loops only
into the dressing mechanism,
the dressed bag propagator in the one-pole approximation
takes the form~\cite{kuk2,kuk5}
\begin{eqnarray}
[G_{DB}]^J_{LL^{\prime}}=\int d{\bf k}\frac{B_L^J({\bf k})
B_{L^{\prime}}^J({\bf k})}{E-E_{\alpha}({\bf k})},
\label{propag}
\end{eqnarray}
where $B_L^J({\bf k})$ is the $\sigma DD$ form factor and 
$E_{\alpha}({\bf k})$ is the total energy of the dressed bag (in the
non-relativistic case $E_{\alpha}({\bf k})\simeq m_{\alpha}+
\varepsilon_{\sigma}({\bf k})$, with $\varepsilon_{\sigma}({\bf k})=
k^2/2m_{\alpha}+\omega_{\sigma}({\bf k})\simeq 
m_{\sigma\alpha}+k^2/2m_{\sigma}$,
and $m_{\sigma\alpha}=m_{\sigma}m_{\alpha}/(m_{\sigma}+m_{\alpha})$ while
$\omega_{\sigma}=\sqrt{m_{\sigma}^2+k^2}$ is the relativistic energy
of $\sigma$-meson). Thus the effective interaction in the $NN$-channel
induced by the intermediate dressed dibaryon production takes the 
form in partial wave representation
\begin{eqnarray}
V_{NqN}=\sum_{JLL^{\prime}}W^J_{LL^{\prime}}({\bf r},{\bf r}^{\prime};E)
\label{vpart}
\end{eqnarray}
with 
\begin{multline}
W^J_{LL^{\prime}}({\bf r},{\bf r}^{\prime};E)=\sum_{M}
\varphi^{JM}_L({\bf r})\lambda^J_{LL^{\prime}}(E)
\varphi^{JM}_{L^{\prime}}({\bf r}^{\prime}),\\
\lambda^J_{LL^{\prime}}(E)=\gamma^2[G_{DB}]^J_{LL^{\prime}},
\label{wpart}
\end{multline}
where $\gamma^2=g_s^4\langle v\rangle ^2$ (see Eqs.~(\ref{cnstr}) 
- (\ref{2s}) for comparison).

It is easy to show~\cite{kuk2,kuk5} that the weight of the $DB$ in the total 
$NN$ wavefunction in the $LL^{\prime}SJ$-channel is proportional to
the energy derivative 
\begin{eqnarray}
\beta^2_{LL^{\prime}J}=-\,
\frac{\partial\lambda^J_{LL^{\prime}}(E)}{\partial E}. 
\label{deriv}
\end{eqnarray}
This derivation is in close analogy to a similar procedure 
for the weight of dressed particle state using the energy derivative 
of the respective polarization operator $\Pi(P^2)$ in the field-theory
formalism. In particular, recently we developed the fully 
covariant EFT-derivation~\cite{kuk6,kuk7} for the relativistic
$NN$-potential at intermediate and short ranges, similar 
to Eqs.~(\ref{propag})-(\ref{deriv}); we also derived in the simplest
one-state approximation a separable form for the 
relativistic $NN$-interaction in channels $^1S_0$ and $^3S_1$-$^3D_1$, 
which fits almost perfectly the respective $NN$ phase shifts for the 
large energy interval 0 - 1000 MeV. 
In contrast to other potentials, 
e.g.  the purely phenomenological Graz separable
$NN$ potential which includes a few dozens adjustable parameters for a lesser
energy interval this high-quality fit has been performed using only four
parameters in the singlet $^1S_0$-channel and a few more for
the coupled-channel case $^3S_1-^3D_1$
%%%%%%%%%%%%%%%%%%%%%%%%%%%%%  Fig. 3 %%%%%%%%%%%%%%%%%%%%%%%%%%%%%%%%%%%%%%
%------------------------------------------------------------------------------
\begin{figure}[hp]\centering
\epsfig{file=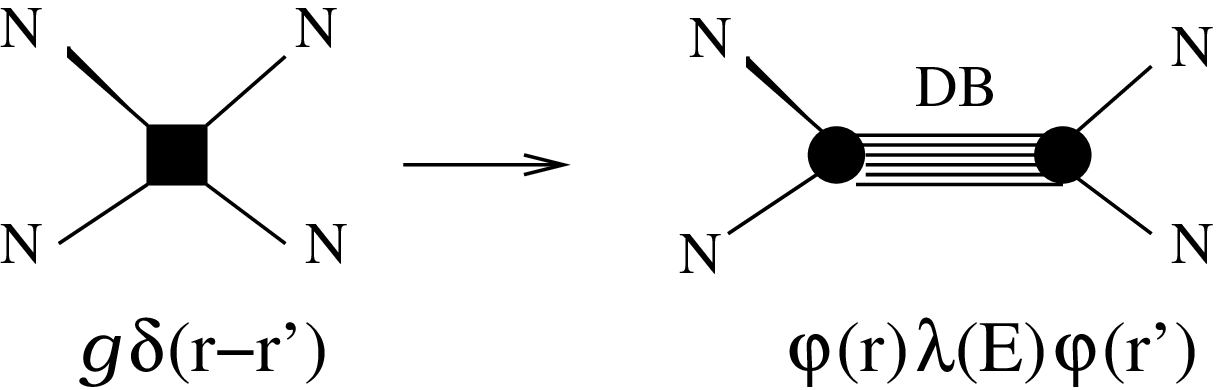,width=0.4\textwidth}
\caption{Different ways for parameterizing the short-range $NN$ 
interaction: in low-energy (pionless) EFT through the four-nucleon
contact term (left) and in the dibaryon model (right) via $s$-channel
intermediate dibaryon.}
\label{f1}
\end{figure}
%------------------------------------------------------------------------------
%%%%%%%%%%%%%%%%%%%%%%%%%%%%%%%%%%%%%%%%%%%%%%%%%%%%%%%%%%%%%%%%%%%%%%%%%%%

The separable form of the short-range $NN$ interaction given in 
Eqs.~(\ref{vpart}) - (\ref{wpart}) can be clearly compared to the contact 
terms in the effective field theory (EFT) approach (pionless) where all the
high-energy physics is parametrized via some contact terms 
(see Fig.~\ref{f1}, the left graph) which cannot be calculated within that 
low-energy approach and must be either 
parametrized somehow phenomenologically or fitted to the 
data~\cite{kuk6,kuk7}.
On the other hand, our short-range mechanism (shown schematically 
in Fig.~\ref{f1}, right) gives just 
the energy dependence for such contact terms.

%%%%%%%%%%%%%%%%%%%%%%%%%%%%%%%%%%%%%%%%%%%%%%%%%%%%%%%%%%%%%%%%%%%%%%
%%********************************************************************
\section{Deuteron structure and dibaryon-induced  short-range 
currents}
\label{sec:deuteron}
%%********************************************************************
%%%%%%%%%%%%%%%%%%%%%%%%%%%%%%%%%%%%%%%%%%%%%%%%%%%%%%%%%%%%%%%%%%%%%%
Similar to the general description of the $NN$-system given 
in Sect.~\ref{sec:dressedbag} 
the total deuteron wavefunction in the DB-model has the form of the 
Fock column Eq.~(\ref{fock}) with at least two components, the $NN$ and the 
dibaryon dressed with $\sigma$-field --- a dressed bag (DB)~\cite{kuk1,kuk2}.
The DB-component $|DB>$ in the second line of Eq.~(\ref{fock}) can be 
presented in the graphic form as a superposition of a bare dibaryon 
(six-quark configuration $s^6[6_X]$) and an infinite series of
$\sigma$-meson loops coupled to the $s^6$-quark core propagator. Thus,
this component includes both the bare and dressed parts. 

Taking further the simple Pade-form (\ref{lame}) of the energy 
dependent factor $\lambda^J_{LL^{\prime}}(E)$ in Eq.~(\ref{wpart}) 
and by calculating the energy derivative (\ref{deriv}) one gets 
easily the weight of the dressed dibaryon component in the
deuteron, which turns out to be $\beta^2\simeq$ 0.025 - 0.035 for
different versions of the model.
{It is very interesting
that this weight of the dibaryon component derived from the energy
dependence of the $\sigma$-loop diagram is rather close to the weight 
of non-nucleonic components (e.g. $\Delta\Delta$ etc.) in the
deuteron deduced within many phenomenological models.}
 Table~\ref{tab:deuteron} gives the 
summary of the static characteristics of the deuteron 
found with this dibaryon model~\cite{kuk2,kask}.

\begin{table*} 
\caption{\label{tab:deuteron} Deuteron properties in different models.} 
\begin{tabular}{llllllll}
\hline
&$E_d$&$P_D$&$r_m$&$Q_d$&$\mu_d$&$A_S$&$D/S$\\
Model&(MeV)&(\%)&(fm)&(fm$^2$)&($\mu_N$)&(fm$^{-1/2}$)&\\[2pt]
\hline
\hline
RSC&2.22461&6.47 $\quad$& 1.957&0.280&0.8429&0.8776&0.0262\\
AV18&2.2245&5.76 & 1.967&0.270&0.8521&0.8850&0.0256\\
Bonn 2001&2.22458&4.85 & 1.966&0.270&0.8521&0.8846&0.0256\\
DB ($NN$ only)&2.22454&5.42 & 2.004&0.286&0.8489&0.9031&0.0259\\
DB ($NN$+6q)&2.22454&5.22 & 1.972&0.275&0.8548&0.8864&0.0264\\[2pt]
Experiment&2.22454(9)$^a$&& 1.971(2)$^b$&0.2859(3)$^c$&0.857406(1)$^d$&
0.8846(4)$^e$&0.0264$^f$\\[2pt]
\hline
\hline
&$^a$Ref.~\cite{a},&&
$^b$Ref.~\cite{b},&
$^c$Ref.~\cite{c} and~\cite{c1},&
$^d$Ref.~\cite{d},&
$^e$Ref.~\cite{e},&
$^f$Ref.~\cite{f}.\\
\end{tabular}
\end{table*}

 The parameters of the model obtained from
fit to the phase shifts of elastic $NN$ scattering in the 
$^3S_1$ - $^3D_1$ channel have the following values
\begin{multline}
\lambda(0)=-385.89\,\, MeV,
\quad E_0=855.29\,\, MeV,\\
\quad a=-0.025,\quad r_0=0.38113\,\, fm,
\quad (b=\sqrt{\frac{3}{2}}r_0).
\label{mtmod}
\end{multline} 
We emphasize here again, that after fitting the phase shifts our approach does
not have any free or adjustable parameters left for the description of the
deuteron properties.  From Table~\ref{tab:deuteron} one can see that the
predictions for the basic deuteron observables found in the above dibaryon
model is generally even in better agreement with respective experimental data
than those for the modern $NN$-potentials calculated conventionally.
  
\begin{figure*}\centering
\epsfig{file=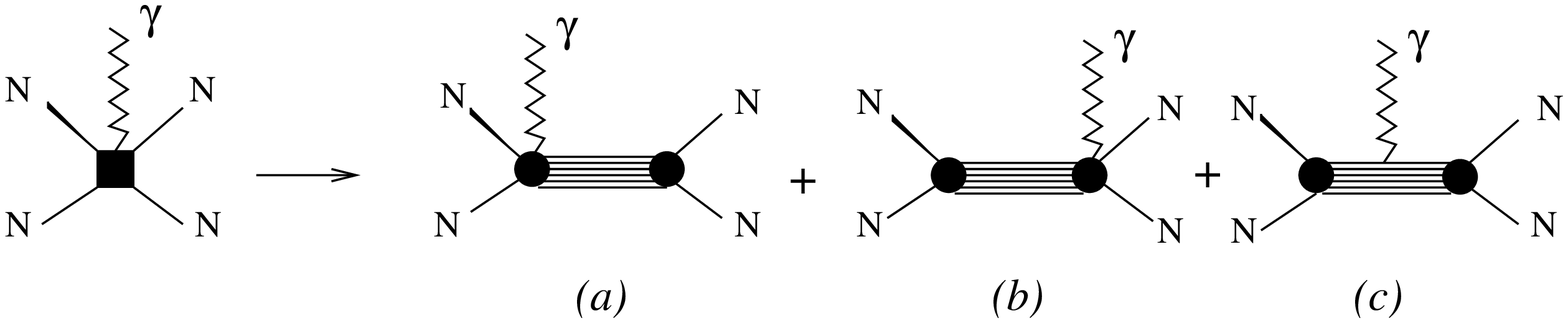,width=0.6\textwidth}
\caption{Schematic representation of the new  electromagnetic
currents induced by intermediate dibaryon generation.}
\label{f2}
\end{figure*}

The modeling of the $NN\to DB$ transition by scalar 
exchanges between quarks makes it possible to consider the ``contact'' 
$\gamma NN\to DB$ vertices (Fig.~\ref{f2})  
in terms of CQM with the minimal electromagnetic interaction of
the constituent quarks, i.e. with the quark current
\begin{eqnarray}
j^{\mu}_{q}(q)&=&\sum_{i=1}^6\hat e_i
F_q(q^2)\bar u(p_i^{\prime})\gamma^{\mu}_iu(p_i),
\label{jq}
\end{eqnarray}
where $q=p^{\prime}-p$,
$\hat e_i= \frac{1}{6}\!+\!\frac{1}{2}\tau^{(i)}_z$ and $F_q(q^2)$
is a form factor of the constituent quark 
%%%%%%%%%%%%%%%%%%%%%%% Fig. 4  %%%%%%%%%%%%%%%%%%%%%%%%%%%%%%%%%%%%%%
%------------------------------------------------------------------------------
%------------------------------------------------------------------------------
%%%%%%%%%%%%%%%%%%%%%%%%%%%%%%%%%%%%%%%%%%%%%%%%%%%%%%%%%%%%%%%%%%%%%%
which can only show itself   
at intermediate momentum transfer $Q^2\gtrsim$ 1 GeV$^2$/c$^2$. It
implies that the constituent quark is an extended object and has obtained its
own electromagnetic form factor, e.g. a monopole one
$F_q(Q^2)=1/(1+Q^2/\Lambda_q^2)$, where the parameter $\Lambda_q$ is expected
to be set by the chiral symmetry scale $\Lambda_{\chi}\simeq4\pi
f_{\pi}\simeq$ 1~GeV.

For definiteness, we consider the diagrams (a) and (c) depicted in 
Fig.~\ref{f2}.
In our model with the scalar exchanges and the CQM current (\ref{jq})
these contact terms are equivalent to the sum of Feynman diagrams
depicted in Fig.~\ref{f5}.
These diagrams describe the two-particle currents in the six-quark 
system. The diagram ($e$) in Fig.~\ref{f5} gives rise to an
additional (i.e. the $\gamma$-induced) contribution to the transition
$s^4p^2\to s^6$ as compared to the mechanism shown in Fig.~\ref{f4}.
Here we demonstrate that within the above model for the short-range $NN$
interaction the minimal quark-photon coupling
$j^{\mu}_{q}(q)A_{\mu}(q)$ 
leads to a non-additive two-nucleon current which does not vanish in the
low-energy limit $q_0, |{\bf q}|\to$ 0. In this limit, only the
contribution of the diagram Fig.~\ref{f5}(e) vanishes because of
orthogonality of the $s^6$ configuration to the quark-cluster states
in the $N$-$N$ channel (i.e. to the configurations $s^4p^2$ and the 
other ones). By contrast, the total contribution of the diagrams
($a$)-($d$) and ($f$) in Fig.~\ref{f5} does not vanish at $q_0\to 0$. 
In each pair of diagrams, ($a$) and ($b$), depicted in Fig.~\ref{f6},
the singular terms $\sim 1/q_0$ are mutually 
canceled, but the remainder, proportional to the momentum {\bf k}
between the $i$-th and  $j$-th quarks and also to the scalar 
$qq$-interaction potential $v_s({\bf k}^2)$, does not vanish in the 
limit $q_0\to 0$ because of non-vanishing the matrix element
$ \langle s^4p^2|v_s|s^6 \rangle $ (see below).

Now we pass to the actual calculations of such diagram contributions.  It
should be stressed here that the current diagrams in Fig.~\ref{f5} 
corresponds just to the `non-diagonal' (transition) electromagnetic 
current which couples two
different channels, i.e. the proper $NN$- and $DB$-channels. These channel
wavefunctions enter the transition matrix elements with a proper own
normalization because any current associated to the $DB$-state is
``normalized'' to the weight of the $DB$-component. Among other things this
makes it possible to avoid any double counting (symmetry properties of quark
configurations are discussed in Appendix~\ref{sec:appenda} which argue
strongly against the repeated contribution in detail).  When the spin part of
the $i$-th quark current (Eq.~\ref{jq}) is taken into account only and a low
energetic $M1$-photon is generated, one can write the following Feynman
amplitudes $M^{\lambda}_{ij}$ for the diagrams depicted in Fig.~\ref{f6}($a$)
and ($b$)
%\begin{widetext}
\begin{multline}
M^{\lambda}_{ij(a)}=\frac{ieg_s^2v_s(k_j^2)}{2m_q}
\bar u(p_i^{\prime})\left\{\hat e_i\sigma_i^{\mu\nu}q_{\nu}
\varepsilon_{\mu}^{(\lambda)*}\frac{2m_q-\!\not k_i+\!\not q}
{2p_i^{\prime}\cdot q}\right.\\
\left.+\frac{2m_q+\!\not k_i-\!\not q}
{-2p_i\cdot q}\hat e_i\sigma_i^{\mu\nu}q_{\nu}
\varepsilon_{\mu}^{(\lambda)*}\right\}u(p_i),\\
M^{\lambda}_{ij(b)}=\frac{ieg_s^2v_s(k_i^2)}{2m_q}
\bar u(p_j^{\prime})\left\{\hat e_j\sigma_j^{\mu\nu}q_{\nu}
\varepsilon_{\mu}^{(\lambda)*}\frac{2m_q-\!\not k_j+\!\not q}
{2p_j^{\prime}\cdot q}\right.\\
\left.+\frac{2m_q+\!\not k_j-\!\not q}
{-2p_j\cdot q}\hat e_j\sigma_j^{\mu\nu}q_{\nu}
\varepsilon_{\mu}^{(\lambda)*}\right\}u(p_j),
\label{mab}
\end{multline}
%\end{widetext}
where $\varepsilon_{\mu}^{(\lambda)*}$ is a space-like photon polarization
vector $\varepsilon^{\mu(\lambda)*}=\{0,{\b\epsilon}^{(\lambda)*}\}$
satisfying the transversality condition 
$\hat{\bf q}\cdot{\b\epsilon}^{(\lambda)*}=$ 0 at $\lambda=\pm$ 1. 

It is easy to verify that the singular terms   
$\sigma_i^{\mu\nu}q_{\nu}\varepsilon_{\mu}^{(\lambda)*}
m_q/(p_i^{\prime}\cdot q)$ and 
$\sigma_i^{\mu\nu}q_{\nu}\varepsilon_{\mu}^{(\lambda)*}
m_q/(\!-\!p_i\cdot q)$ cancel each other in sum 
$M^{\lambda}_{ij(a)}+M^{\lambda}_{ij(b)}$
in the limit $q_0\to 0$. As a result, we obtain
from Eq.~(\ref{mab}) in the non-relativistic
approximation $q_0/m_q\ll$ 1 the three-dimensional operator
\begin{widetext}
\begin{multline}
V_{Nq\gamma}=\frac{ieg_s^2}{2m_q}\sum_{i\!=\!1}^3\sum_{j\!=\!4}^6
{\b\epsilon}^{(\lambda)*}\cdot\left\{v_s({\bf k}^2_{j})
\left(\frac{\hat{\bf q}\cdot{\bf k}_i}{m_q}
[{\b\sigma}_i\times\hat{\bf q}]-
\frac{[{\b\sigma}_i\times{\bf k}_i]}{m_q}\right)
+v_s({\bf k}^2_{i})
\left(-\,\frac{\hat{\bf q}\cdot{\bf k}_j}{m_q}
[{\b\sigma}_j\times\hat{\bf q}]-
\frac{[{\b\sigma}_j\times{\bf k}_j]}{m_q}\right)\right\}\\
\times (2\pi)^3\delta^3({\bf p}_i+{\bf p}_j-
{\bf p}^{\prime}_i-{\bf p}^{\prime}_j-{\bf q})
\label{vngq}
\end{multline}
%\end{widetext}
%%%%%%%%%%%%%%%%%%%%%%% Fig. 5 %%%%%%%%%%%%%%%%%%%%%%%%%%%%%%%%%%%%%%
%------------------------------------------------------------------------------
\begin{figure*}[hp]\centering
\epsfig{file=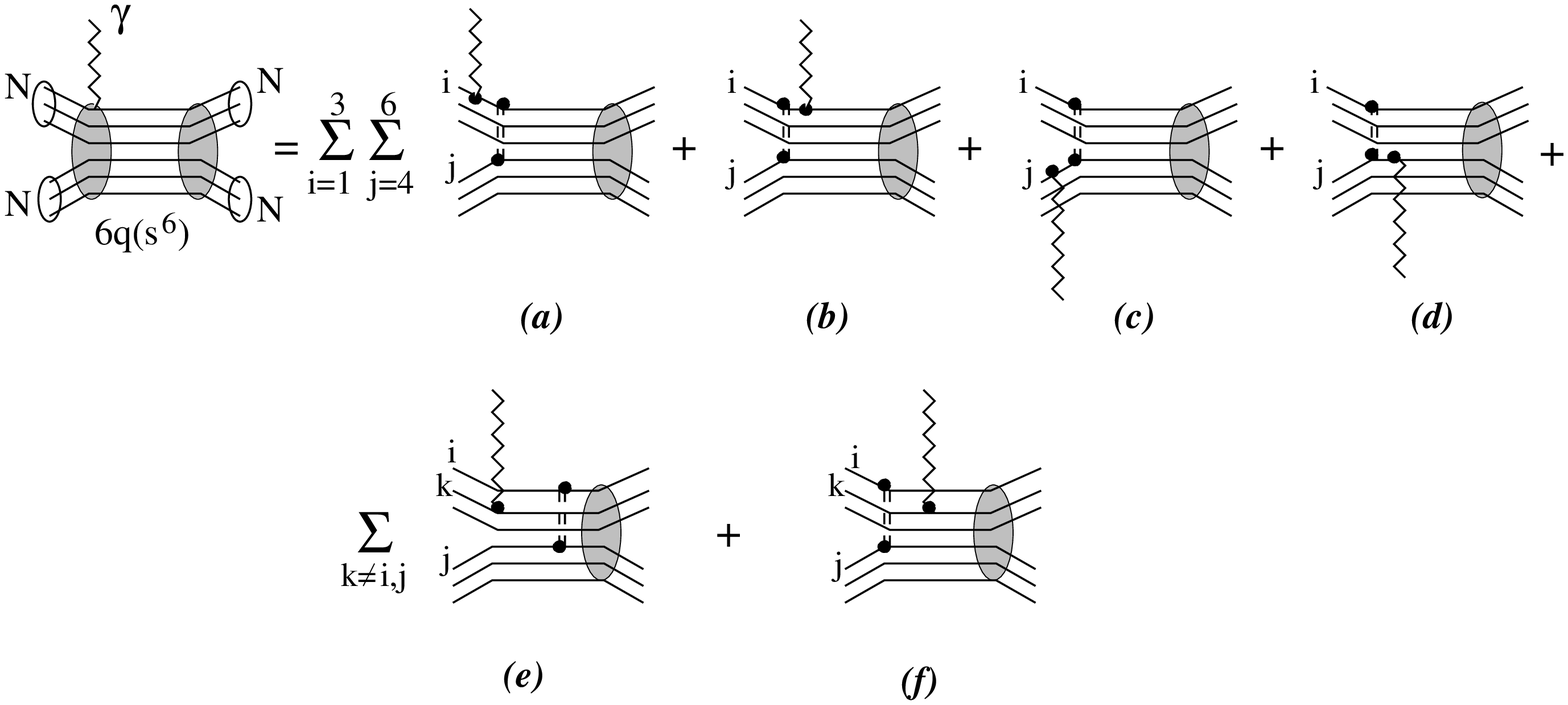,width=0.8\textwidth}
\caption{The diagram series illustrating the contact $V_{Nq\gamma}$ terms
contributing to the ``non-diagonal'' current.}
\label{f5}
\end{figure*}
%------------------------------------------------------------------------------
%%%%%%%%%%%%%%%%%%%%%%%%%%%%%%%%%%%%%%%%%%%%%%%%%%%%%%%%%%%%%%%%%%%%%
\end{widetext}
defined on non-relativistic quark wavefunctions of the CQM. This
operator describes the transition from the $NN$ channel to the 
$6q$-bag with emission of a $M1$ $\gamma$-quantum, i.e. a ``contact''
$NN\to DB+\gamma$ interaction, schematically shown in 
Fig.~\ref{f2}(a).

Now we can calculate the effective contact vertex 
$NN\Leftrightarrow NN\gamma$ 
(see Fig.~\ref{f2}a) on the basis of the quark operator $V_{Nq\gamma}$ 
in terms of the quark-microscopic version of the $DB$ model.
Recall that in our model the diagrams in Fig.~\ref{f6} taken without
electromagnetic insertions are simply the pairwise $q$-$q$ interaction: 
\begin{multline}
V_{Nq}=g_s^2\sum_{i\!=\!1}^3\sum_{i\!=\!4}^6[v_s({\bf k}^2_{i})
+v_s({\bf k}^2_{j})]\\
\times(2\pi)^3\delta^3({\bf p}_i+{\bf p}_j-
{\bf p}^{\prime}_i-{\bf p}^{\prime}_j-{\bf q}),
\label{vnq}
\end{multline}
which describes the transition from the $NN$- to the $6q$-bag
channel (see Fig.~\ref{f4}).
This observation points toward the proper solution of the problem of contact
$NN\Leftrightarrow NN\gamma$ (further on we use 
the notation ``$NqN\gamma$'' for brevity)
interaction in our approach. Namely,  we calculate the non-local 
$NqN\gamma$-interaction operator in the $NN$ Hilbert space 
$V_{NqN\gamma}(r,r^{\prime})$ by the same way as the
non-local $NqN$-interaction operator $V_{NqN}(r,r^{\prime})$ in 
Eqs.~(\ref{cnstr}) -(\ref{2s}). 
%%%%%%%%%%%%%%%%%%%%%%% Fig. 6 %%%%%%%%%%%%%%%%%%%%%%%%%%%%%%%%%%%%%%
%------------------------------------------------------------------------------
\begin{figure*}\centering
\epsfig{file=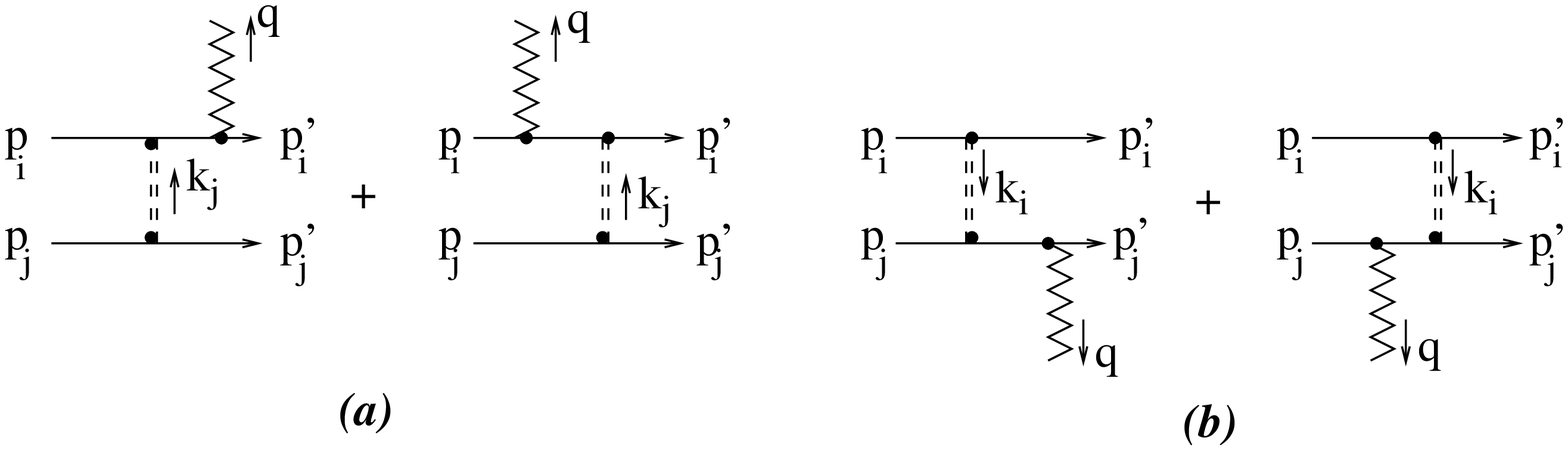,width=0.7\textwidth}
\caption{The part of diagrams shown in Fig.~\ref{f5} contributing
to the non-diagonal dibaryon current with explicit notations of the
kinematic variables. The double-dashed line denotes the scalar
exchange.}
\label{f6}
\end{figure*}
%------------------------------------------------------------------------------
%%%%%%%%%%%%%%%%%%%%%%%%%%%%%%%%%%%%%%%%%%%%%%%%%%%%%%%%%%%%%%%%%%%%%

We obtain finally (see Appendix B for detail)
the $NqN\gamma$ (contact) term searched for
(as the sum of two graphs, ($a$) and ($b$), in Fig.~\ref{f2})
\begin{widetext}
\begin{multline}
V_{NqN\gamma}^{(\lambda)}({\bf q};r,r^{\prime})=
\frac{eZ}{2M_N}\left\{i\left[
\frac{{\b\sigma}_p+{\b\sigma}_n}{2}\times{\bf q}
\right]\cdot{\b\epsilon}^{(\lambda)*}G^S_M(q^2)
+i\left[\frac{{\b\sigma}_p-{\b\sigma}_n}{2}\times{\bf q}
\right]\cdot{\b\epsilon}^{(\lambda)*}G^V_M(q^2)\right\}\\
\times\left\{\frac{1}{q}j_1(qr/2)\frac{d\varphi_{2S}(r)}{dr}
\frac{\lambda(E^{\prime})}{2M_N}\varphi_{2S}(r^{\prime})
+\varphi_{2S}(r)\frac{\lambda(E)}{2M_N}
\frac{1}{q}j_1(qr^{\prime}/2)
\frac{d\varphi_{2S}(r^{\prime})}{dr^{\prime}}\right\},
\label{fin}
\end{multline}
\end{widetext}
where $G^S_M(0)=\mu_p+\mu_n$ and $G^V_M(0)=\mu_p-\mu_n$ (the origin of the
nucleon form factors  $G^S_M$ and $G^V$ in the quark model results of
Eq.~(\ref{fin}) type is discussed in Appendix C). Our basic expression 
for the transition dibaryon current still does not take into account
possible effects which must affect the predictions of our model
(viz. the relativistic effects and quark boost contributions 
which should be essential at $Q^2\sim$ 1 GeV$^2$/c$^2$~\cite{phil},
and other contact terms with inclusion of pseudo-scalar and vector-meson
exchanges~\cite{phil,ris} etc.). So, to account of all these effects
we renormalize our contact $NN\Leftrightarrow NN\gamma$ 
vertex using some renormalization 
factor $Z$ in Eq.~(\ref{fin}). It is felt that the value
$Z\approx$ 1 $\pm$ 0.3 is reasonable since a precision of 
10 - 30\% is typical for standard quark model evaluations of the hadron
magnetic moments. We show below that when choosing a reasonable value 
for a single free constant $Z=$ 0.7 the contact 
term given in Eq.~(\ref{fin}) leads to a considerable improvement 
in description of isoscalar magnetic properties of the deuteron.

One can get a general expression for the electromagnetic current in 
the $NN$ system, and also in the deuteron, starting with the quark
current (\ref{jq}) which has already been used for finding the
contact $NqN\gamma$ vertex (\ref{fin}). When the total deuteron 
wavefunction in the Fock-column form  Eq.~(\ref{fock}) is considered, 
the diagonal matrix element of the quark current (\ref{jq}) 
can be represented as
\begin{widetext}
\begin{multline}
 \langle d|\sum_{i\!=\!1}^6j^{\mu}_i(q)
\varepsilon_{\mu}^{(\lambda)}(q)|d \rangle\,=
cos^2\theta_{{\scriptscriptstyle N}q}
\langle d(NN)|J_{\scriptscriptstyle N}^{\mu}\varepsilon_{\mu}^{(\lambda)}
+V_{NqN\gamma}^{(\lambda)}|d(NN) \rangle\\
+sin^2\theta_{{\scriptscriptstyle N}q}
\langle DB|\sum_{i\!=\!1}^6j^{\mu}_i\varepsilon_{\mu}^{(\lambda)}
|DB \rangle+2cos\,\theta_{{\scriptscriptstyle N}q}
sin\,\theta_{{\scriptscriptstyle N}q}
 \langle d(NN)|\sum_{i\!=\!1}^6j^{\mu}_i\varepsilon_{\mu}^{(\lambda)}
|DB \rangle .
\label{basic1}
\end{multline}
\end{widetext}
The last two terms in Eq.~(\ref{basic1}) are nothing else but 
contributions of the graphs shown in Figs.~\ref{f5}(f) and (e)
respectively.

It is worth to summarize here our main findings. Within our 
two-component $NN$ interaction model, the minimal substitution
leads basically to two different two-particle currents (in addition 
to the single-nucleon current  $J_{{\scriptscriptstyle N}}^{\mu}$ 
written in the first term of Eq.~(\ref{basic1})), viz. the 
transition ($NN\to DB$) contact term as given by Eq.~(\ref{fin}) 
and the standard six-quark bag current given by the second and 
third terms in Eq.~(\ref{basic1}).

In the nucleon sector we further replace the quark-model current
of the nucleon $J_{{\scriptscriptstyle N}QM}^{\mu}$ with the standard 
representation of $J_{{\scriptscriptstyle N}}^{\mu}$ in terms of
the phenomenological form factors
given in Appendix C by Eqs.~(\ref{ejn}) and (\ref{nfqm}). However, 
the two-body current in the last two terms of Eq.~(\ref{basic1})
(which gives only a small correction to the
single-nucleon current $J_{{\scriptscriptstyle N}}^{\mu}$) is calculated
here on the basis of the constituent quark model (see Appendix C 
for details).

%%%%%%%%%%%%%%%%%%%%%%%%%%%%%%%%%%%%%%%%%%%%%%%%%%%%%%%%%%%%%%%%%%%%%%     
\section{Isoscalar M1 and E2 transition amplitudes}
\label{sec:transampl}
%%%%%%%%%%%%%%%%%%%%%%%%%%%%%%%%%%%%%%%%%%%%%%%%%%%%%%%%%%%%%%%%%%%%%%

The effective electromagnetic operator of the isoscalar current 
$V_{NqN\gamma}^{(\lambda)}$ derived in the previous section is defined, 
by construction, in a Hilbert space of the $NN$ component of the whole
two-component system. Thus, it should 
be bracketed between the initial and final states just in the 
$NN$-channel. So, we look here at the application of this new current
operator to the three observables: ({$i$}) radiative capture 
$\vec n+p\to d+\vec\gamma$ of spin-polarized neutrons by hydrogen;
($ii$) the deuteron magnetic form factor $B(Q^2)$ in the region of
its diffraction minimum; and ($iii$) the very tiny correction to the 
magnetic moment of deuteron.

In the radiative capture process, our main interest lies in the calculation 
of the circular polarization of $\gamma$-quanta emitted after capture at 
thermal energies. It includes both M1 and E2 isoscalar transitions. We 
can contrast for this process the ``contact'' isovector and isoscalar 
transitions, where the corresponding $\pi$-exchange term 
has a long range and corresponds to the isovector transition while
the scalar-exchange term relates to the short-range $\sigma$-, 2$\pi$- or 
glueball-exchange between quarks in both nucleons (or to the 
instanton-induced interaction as well) and corresponds to the isoscalar 
transition. The above $\pi$-meson isovector current
contributes to the total isovector amplitude for the $^1S_0\to ^3S_1$  
transition, which is generally large, and thus this term does not
affect strongly the $P_{\gamma}$-value which is governed 
just by an interference
between isovector M1 and isoscalar M1+E2 amplitudes. The main point here 
is that the single-nucleon isoscalar transition is strongly
suppressed due to orthogonality of the initial and final radial 
wavefunctions. In this case, the small isoscalar contribution to the 
$P_{\gamma}$ and to an angular asymmetry of the photons can be of
crucial importance due to their interference with the large isovector 
amplitude~\cite{rho4,che,park2}. The isoscalar M1 current can be 
also very important 
for the deuteron magnetic form factor $B(Q^2)$ in the area where the
contribution of single-nucleon current almost vanishes. So, this new
isoscalar current can affect essentially the behavior of $B(Q^2)$ near
its minimum.

To fix uniquely the relative signs of partial transition amplitudes
(and for a meaningful comparison between predictions of different 
models) we use in all our calculations a common expansion of the
photon plane wave into vector spherical harmonics (see, e.g. 
Refs.~\cite{akh,gra}) and a standard choice~\cite{ed} for the phase
factors of Clebsch-Gordon coefficients and spherical functions. This 
choice fixes the sign of the $E2$ amplitude uniquely. The problem with the
relative sign of $E2$ amplitude would arise if one calculates the
$E2$ and $M1$ amplitudes separately  
(see, e.g., the detailed discussion of the $E2$-sign problem in 
Refs.~\cite{khrip,shma}).

This general formalism, common for two different electromagnetic 
processes, has been used in the present work jointly with our new 
$NN$-force model to estimate a non-additive two-body current 
contribution.

\subsection{General consideration }

Let us start here with the single-nucleon current. The expansion of the
circularly polarized $\gamma$-quanta emission (with $\lambda=\pm$ 1)
operator into electric  and magnetic  multipoles takes the
form~\cite{akh,gra} 
\begin{widetext}
\begin{multline}
-\vec J_N\cdot\vec\epsilon^{\,(\lambda)*}(\hat q)\,
e^{-i\vec q\cdot\vec r/2}=\frac{\lambda}{\sqrt{2}}\sum_{l=1}^{\infty}
\sqrt{4\pi(2l+1)}(-i)^l\,\vec J_N\cdot\left\{j_l(qr/2)
\vec Y^*_{l,l,\lambda}(\hat r)\right.\\
\left.-i\lambda\left[\sqrt{\frac{l\!+\!1}{2l\!+\!1}}j_{l\!-\!1}(qr/2)
\vec Y^*_{l,l\!-\!1,\lambda}(\hat r)-\sqrt{\frac{l}{2l\!+\!1}}
j_{l\!+\!1}(qr/2)\vec Y^*_{l,l\!+\!1,\lambda}(\hat r)\right]
\right\}
\label{mult}
\end{multline}
\end{widetext}
in which we employ for vector spherical harmonics 
\begin{eqnarray}
\vec Y_{j,l,\lambda}(\hat r)=\sum_{\varkappa}
(l(\lambda\!-\!\varkappa)1\varkappa|j\lambda)Y_{l,\lambda\!-\!\varkappa}
(\hat r)\,\vec\epsilon^{\,(\varkappa)}(\hat q)).
\label{vg}
\end{eqnarray}
with the spherical polarizations
\begin{eqnarray}
\vec\epsilon^{\,(\pm1)}(\hat q)=\mp\frac{\hat x\pm\hat y}{\sqrt{2}},
\qquad \vec\epsilon^{\,(0)}(\hat q)=\hat q,
\label{eps}
\end{eqnarray}
which are determined in the reference frame related to the photon
with the quantization axis $Z_0$ directed along the photon momentum
${\bf q}$, $\hat{\bf z}_0=\hat{\bf q}$.

At thermal energies of neutrons, which corresponds to the 
velocity $v=$ 2200 m/s,
one can neglect the electric dipole transition as the initial $P$-wave
is strongly depressed. In Eq.~(\ref{mult}) only two
lowest multipoles, M1 and E2, remain, because the initial $^3D_1$-wave is
admixtured to the basic $^3S_1$-wave by the strong tensor force;
we emphasize here that our model includes also a significant
short-range tensor force originated from the intermediate dibaryon in
addition to the conventional OPE tensor force. 
The operator of single-nucleon transverse current
\begin{multline}
e\vec J^{\,T}_N=\frac{1}{2m_N}\left\{\frac{e_p+e_n}{2}
(-2i\vec\nabla^{\,T})\right.\\
\left.+e[(\mu_p\vec\sigma_p+\mu_n\vec\sigma_n)\times2\vec\nabla_r]\right\}
\label{jn}
\end{multline}
contains the $\vec\nabla^{\,T}$, a transverse component of a gradient which
operates on the initial $np$ wavefunction and a gradient
$\vec\nabla_r=\frac{1}{2}\vec\nabla_{r/2}$ operating only on the plane wave
$e^{-i\vec q\cdot\vec r/2}$, associated with the momentum of emitted photon
$\vec q$. Inserting $e\vec{J}^{\,T}_N$ into Eq.~(\ref{mult}) one gets, after
some algebra, the following representation for the transition amplitude $np\to
d\gamma$ in an arbitrary coordinate frame XYZ
\begin{multline}
T_{MM^{\prime}}^{(\lambda^{\prime})}(\varphi,\theta)=\sum_{\lambda=\pm1}
\left[M1_{MM^{\prime}}^{(\lambda)}
{\cal D}^{(1)}_{\lambda\lambda^{\prime}}(\varphi,\theta,0)\right.\\
\left.+E2_{MM^{\prime}}^{(\lambda)}
{\cal D}^{(2)}_{\lambda\lambda^{\prime}}(\varphi,\theta,0)\right].
\label{tm}
\end{multline}
Here the photon emission angles are given in the reference frame XYZ 
where the axis Z is chosen conveniently along the polarization
vector of incident neutron. Then the quantum numbers 
$MM^{\prime}\lambda^{\prime}$ are projections onto the 
quantization axis Z of the initial and final spin of the $np$ system
and of the photon total angular momentum $l$ respectively, while the
photon helicity $\lambda=\pm$1 is defined, as above, in its own
reference frame X$_0$Y$_0$Z$_0$. In correspondence with this 
definition of the reference frame in Eqs.~(\ref{mult}) and (\ref{tm})
the matrix elements for the M1- and E2-transitions are calculated
for the fixed values $\lambda=\pm$1 of the photon helicity, but in
an arbitrary reference frame XYZ in which the initial-state 
wavefunctions of $np$ scattering are given
\begin{widetext}
\begin{multline}
\Psi^{np}_M(\vec r,\vec p_{n})=\frac{1}{r}{^{\,1\!}S}_0(r,p_{n})Y_{00}
(\hat r)\sum_{\lambda_p,\lambda_n}({\scriptstyle\frac{1}{2}}\lambda_p
{\scriptstyle\frac{1}{2}}\lambda_n|00)\chi_{{\scriptscriptstyle\lambda}_p}
\chi_{{\scriptscriptstyle\lambda}_n}\delta_{\scriptscriptstyle M,0}
+\frac{1}{r}{^{\,3\!}S}_1(r,p_{n}){\cal Y}^{01}_{1M}(\hat r)+
\frac{1}{r}{^{\,3\!}D}_1(r,p_{n}){\cal Y}^{21}_{1M}(\hat r)
\label{npi}
\end{multline}
\end{widetext}
and the final-state wavefunction reads
\begin{eqnarray}
\Psi^d_{M^{\prime}}(\vec r)=\frac{u(r)}{r}{\cal Y}^{01}_{1M}(\hat r)
+\frac{w(r)}{r}{\cal Y}^{21}_{1M}(\hat r),
\label{df}
\end{eqnarray}
where ${\cal Y}^{LS}_{JM}(\hat r)=
\sum\limits_{\scriptscriptstyle M_L,\,M_S}(LM_LSM_S|JM)
Y_{LM_L}(\hat r)\chi_{\scriptscriptstyle SM_S}$. The states 
${^1S}_0(r,p_{n})$, ${^3S}_1(r,p_{n})$ and ${^3D}_1(r,p_{n})$ 
are fixed by the dynamics of the $np$-interaction, i.e. the continuum 
$S$-waves are normalized by the respective scattering lengths
\begin{multline}
{^1S}_0(r,p_{n})\to r-a_t,\qquad {^3S}_1(r,p_{n})\to r-a_s,\\
p_{n}\to 0,
\label{sw}
\end{multline}
while the ${^3D}_1(r,p_{n})$-wave normalization is fixed by the 
$^3S_1-^3D_1$ tensor mixing in the initial state.

After elementary but lengthy calculations, one gets the following 
formulas for the matrix elements at the right hand side of 
Eq.~(\ref{tm})
\begin{widetext}
\begin{multline}
M1^{(\lambda)}_{MM^{\prime}}=\int d^3r\Psi^{d*}_{M^{\prime}}(\vec r)
\frac{ieq}{2m_N}\,\lambda\left\{(\mu_n\!-\!\mu_p)j_0(qr/2)\,
\vec\epsilon^{\,(\lambda)*}\cdot\frac{\vec\sigma_p\!-\!\vec\sigma_n}{2}
\right.\\
\left.+(\mu_n\!+\!\mu_p)\left[j_0(qr/2)\,\vec\epsilon^{\,(\lambda)*}\cdot
\frac{\vec\sigma_p+\vec\sigma_n}{2}-
\sqrt{\frac{1}{2}}j_2(qr/2)\sqrt{4\pi}\,\vec Y^*_{1,2,\lambda}
(\hat r)\cdot\frac{\vec\sigma_p+\vec\sigma_n}{2}\right]
\right.\\
\left.+\frac{1}{2}(j_0(qr/2)+j_2(qr/2))\,\vec\epsilon^{\,(\lambda)*}
\cdot\frac{1}{2}\vec L\right\}\Psi^{np}_M(\vec r,\vec p_{n}).
\label{m1}
\end{multline}
\end{widetext}
The first term in the curly brackets corresponds to the isovector
M1 transition $^1S_0(NN)\to d(^3S_1)$ while the remaining two terms
describe the isoscalar transitions in the coupled $^3S_1$ - $^3D_1$ channels 
 generated by the spin-dependent and convection currents respectively. 
In contrast to this, the E2-transition amplitude is purely isoscalar,
although it consists of two terms, convection (first) and spin-dependent 
(second) ones, similarly to the M1 transition, 
\begin{widetext}
\begin{multline}
E2^{(\lambda)}_{MM^{\prime}}=\int d^3r\Psi^{d*}_{M^{\prime}}(\vec r)
\frac{ieq}{2m_N}\,\left\{\frac{\lambda}{2}[j_0(qr/2)+j_2(qr/2)]
\,\vec\epsilon^{\,(\lambda)*}\cdot\frac{1}{2}\vec L\right.\\
\left.-\sqrt{\frac{5}{2}}(\mu_n\!+\!\mu_p)
j_2(qr/2)\sqrt{4\pi}\,\vec Y^*_{2,2,\lambda}(\hat r)\cdot
\frac{\vec\sigma_p+\vec\sigma_n}{2}
\right\}\Psi^{np}_M(\vec r,\vec p_{n}).
\label{e2}
\end{multline}
\end{widetext}

The total $np\to d\gamma$ reaction cross section for unpolarized 
neutrons can be expressed through the respective amplitudes 
(\ref{m1}) - (\ref{e2}) in the following way
\begin{multline}
\sigma^{tot}_{unpol}=\frac{m_n}{p_n}\alpha|\vec q|
\frac{\vec q^{\,2}}{4m_N^2}\frac{1}{3}\sum_{MM^{\prime}}
\sum_{\lambda=\pm1}4\pi\left[|M1^{(\lambda)}_{MM^{\prime},I=1}|^2\right.\\
\left.+|M1^{(\lambda)}_{MM^{\prime},I=0}|^2+
|E2^{(\lambda)}_{MM^{\prime}}|^2\right]
\label{stot}
\end{multline}
For the further calculations one can use the well-known 
properties of the Wigner ${\cal D}$-functions,
which gives actually the angular behavior of the interference
term between M1 and E2 amplitudes. For example, in radiative 
capture of spin-polarized neutrons by  spin-polarized protons,
$\vec n+\vec p\to d+\gamma$, 
the angular anisotropy for emission of $\gamma$-quanta
in respect to the spin polarization axis of the initial nucleons
can be described.

Let us consider now the asymmetry in circular 
polarization of $\gamma$-quanta on the basis of Eqs.~(\ref{tm}) and 
(\ref{m1}) - (\ref{stot}), as measured in the experiment~\cite{lob}.
The differential cross section for circularly-polarized
$\gamma$-quanta emission in forward direction (i.e. along the 
spin polarization $\lambda_n$ of incident neutron) can be written
in terms of the helicity amplitudes (\ref{m1}) - (\ref{e2})
\begin{widetext}
\begin{equation}
\sigma_{\lambda}(\lambda_n)=\frac{m_n}{p_n}\alpha|\vec q|
\frac{\vec q^{\,2}}{4m_N^2}\frac{1}{2}
\sum_{\lambda_p}\sum_{M^{\prime}}\left\vert\sum_{M}
({\scriptstyle\frac{1}{2}}\lambda_p{\scriptstyle\frac{1}{2}}\lambda_n|00)
M1^{(\lambda)}_{MM^{\prime},I=1}
+({\scriptstyle\frac{1}{2}}\lambda_p
{\scriptstyle\frac{1}{2}}\lambda_n|1M)
\left[M1^{(\lambda)}_{MM^{\prime},I=0}+E2^{(\lambda)}_{MM^{\prime}}
\right]\right\vert^2.
\label{siglam}
\end{equation}
\end{widetext}
For sake of brevity, the differential cross section 
$d\sigma(\lambda_n,\theta\!=\!0)/d\Omega$ for $\gamma$-quanta 
emission at zero angle in respect
to the neutron polarization vector in  
case when the neutron spin projection onto the quantization axis 
equals to $\lambda_n$ is denoted as 
$\sigma_{\lambda}(\lambda_n)$. In Eq.~(\ref{siglam}) we have omitted the
Wigner ${\cal D}$-functions depicted in Eq.~(\ref{tm}) because the 
respective sums over $\lambda^{\prime}$ are reduced, at $\theta=$ 0, 
to a trivial factor 1.

The differential cross section 
$d\sigma(\lambda_n,\lambda_p;\theta)/d\Omega$ for the photon 
emission into an arbitrary angle $\theta$ can be found using 
the simple transformation of Eq.~(\ref{siglam}) by replacing the sum
$\frac{1}{2}\sum_{\lambda_p}$ by $\sum_{\lambda\!=\!\pm1}$
and by replacing the Wigner ${\cal D}$-functions on the r.h.s. of 
Eq.~(\ref{siglam}).

\subsection{The $\vec n+p\to d+\vec\gamma$ reaction}

Using the general formulas~(\ref{stot}) and  (\ref{siglam}) for 
the helicity dependent cross sections 
one can find the circular polarization $P_{\gamma}(\lambda_n)$ and 
angular anisotropy $\eta$ for the fixed initial values of $\lambda_n$
(or $\lambda_n$,$\lambda_p$)
\begin{multline}
P_{\gamma}(\lambda_n)=\frac{\sigma_{\lambda=1}(\lambda_n)
-\sigma_{\lambda=-1}(\lambda_n)}{\sigma_{unpol}}\\
=\frac{\sum\limits_{\lambda=\pm1}\lambda\sigma_{\lambda}(\lambda_n)}
{\frac{1}{2}\sum\limits_{\lambda_n}
\sum\limits_{\lambda=\pm1}\sigma_{\lambda}(\lambda_n)}
\label{pgam}
\end{multline}
\begin{multline}
\eta(\lambda_n,\lambda_p)=\\
\frac{d\sigma(\lambda_n,\lambda_p,\theta\!=\!\frac{\pi}{2})/d\Omega
-d\sigma(\lambda_n,\lambda_p,\theta\!=\!0)/d\Omega}
{d\sigma(\lambda_n,\lambda_p,\theta\!=\!\frac{\pi}{2})/d\Omega
+d\sigma(\lambda_n,\lambda_p,\theta\!=\!0)/d\Omega}
\label{eta}
\end{multline}

The $M$1 and $E$2 amplitudes which contribute to the cross sections  
(\ref{stot}) and (\ref{siglam}) are given in Appendix D in their
exact form. It is important to stress, that the dependence of the 
M1- and E2-transition matrix elements upon the momentum transfer $q$ 
in Eqs.~(\ref{m1s1}) - (\ref{i0i2}) is rather weak at low energies and
becomes quite significant only for $e$-$d$ scattering in the region of
moderate and high momenta transfer (see below). This means that when
applying formulas~(\ref{m1s1}) - (\ref{i0i2}) to the $np\to d\gamma$
cross section at the thermal energies the integrals $I_2$ can be
neglected while $j_0(qr/2)$ in the integrand of $I_0$ can be replaced
by unity.

As a result, eventually the expression for $P_{\gamma}$ can be presented
via more simple ``reduced'' matrix elements  
\begin{multline}
M1_{I\!=\!1}=(\mu_p-\mu_n)I_0(u,{}^1S_0),\\
M1_{I\!=\!0}=(\mu_p+\mu_n)\left[I_0(u,{}^3S_1)-\frac{1}{2}I_0(w,{}^3D_1)
\right]\\
+\frac{3}{8}I_0(w,{}^3D_1),\\
E2_{I\!=\!0}=\frac{3}{8}I_0(w,{}^3D_1).
\label{mpg}
\end{multline}
with $I_0(f,Z)=\int_0^{\infty}f(r)Z(r,p_n)dr$, where 
$Z(r,p_n)$ can be any of the scattering wavefunctions in $^1S_0$,
$^3S_1$ or $^3D_1$ channels. Thus we get eventually
\begin{multline}
P_{\gamma}(\lambda_n)=(-1)^{1/2+\lambda_n}P_{\gamma},\\
P_{\gamma}=2Re\left\{\frac{M1_0}{M1_1}+\frac{E2_0}{M1_1}\right\},
\label{pgamma}
\end{multline}
in which the factor $(-1)^{1/2\!+\!\lambda_n}$ in front of $P_{\gamma}$
reflects only that 
dependence on $\lambda_n$, which is deduced from the Clebsch-Gordon
coefficient $({\scriptscriptstyle\frac{1}{2}}\lambda_p
{\scriptscriptstyle\frac{1}{2}}\lambda_n|00)=(-1)^{1/2\!+\!\lambda_n}
{\scriptscriptstyle\sqrt{\frac{1}{2}}}$ in the first term of 
Eq.~(\ref{siglam}). Moreover, since in the limit $p_n\to$ 0 all
ratios of the matrix elements in Eqs.~(\ref{m1s1}) and (\ref{mpg}) become
real, the symbol $Re$ can be omitted here.

Quite similar considerations of the angular anisotropy $\eta$ yield a
formula quadratic with respect to the matrix elements in Eqs.~(\ref{mpg})
and bilinear on spin-polarizations of neutron and 
proton~\cite{rho4,che}. 

In the literature there are calculations for $P_{\gamma}$ with the RSC
potential, but published results~\cite{khrip} do not include any details and
any patterns due to the interference of various $M1$- and $E2$-terms.
Therefore, we make also a parallel calculation for the value of $P_{\gamma}$
with the well-known Reid $NN$ potential (version Reid 93~\cite{e}).  Thus, the
detailed comparison for all partial contributions between our and the
conventional RSC-potential model sheds light on the delicate balance of
different isoscalar current components to the total value of $P_{\gamma}$.

\subsection{Deuteron magnetic form factor}

Usually the deuteron magnetic form factor includes a contribution of the
transverse current~(Eq.\ref{jn}) as a whole without explicit separation into
$M1$ and $E2$ multipoles. However, in calculations of the deuteron magnetic
form factor we still can employ the helicity amplitudes (\ref{m1}) and
(\ref{e2}) derived above by summing and substituting the deuteron
wavefunctions $u$ and $w$ instead of the wavefunctions $^3S_1$ and $^3D_1$ in
the continuum. In this substitution the isovector part (the first term of
Eq.~(\ref{m1})) automatically vanishes while in the isoscalar part the
electric charge $e$ (in the convection current term) and the magnetic moment
($\mu_p+\mu_n$) in the spin-dependent term are replaced with their respective
isoscalar counterparts, viz. the isoscalar electric and magnetic form factors
of nucleon
\begin{multline}
G^s_E(q^2)=G^p_E(q^2)+G^n_E(q^2),\\
G^s_M(q^2)=G^p_M(q^2)+G^n_M(q^2),\quad
q^2=q^2_0-\vec q^{\,2}.
\label{gsgm}
\end{multline}
As a result, the sum of M1- and E2 contributions (see Eqs.~(\ref{m1s0}) - 
(\ref{e2c}) in Appendix D) is transformed into the well known formula for 
the deuteron magnetic form factor
\begin{widetext}
\begin{multline}
G^d_M(q^2)=\sqrt{\frac{2}{3}}\frac{\sqrt{-q^2}}{2m_N}\left\{\frac{3}{4}
G^s_E(q^2)\int_{0}^{\infty}w^2(r)[j_0(qr/2)+j_2(qr/2)]dr\right.\\
\left.+G^s_M(q^2)\int_{0}^{\infty}
\left[\left(u^2(r)-\frac{1}{2}w^2(r)\right)
j_0(qr/2)+\frac{w(r)}{\sqrt{2}}\left(u(r)+\frac{w(r)}{\sqrt{2}}\right)
j_2(qr/2)\right]dr\right\},
\label{gdm}
\end{multline}
\end{widetext}
where the factor $\sqrt{\frac{2}{3}}$  accounts for the
averaging of the amplitude squared over the spin projections. This  
gives the standard nor\-ma\-li\-za\-tion of the deuteron magnetic form factor 
which leads to the conventional expression for 
the deuteron structure fun\-c\-ti\-ons $A$ and $B$ 
\begin{multline}
A(q^2)=\left [G^d_{C0}(q^2)\right ]^2+\left [G^d_{C2}(q^2)\right ]^2
+\left [G^d_M(q^2)\right ]^2,\\
B(q^2)=2(1+\eta_d)\left [G^d_M(q^2)\right ]^2,\quad 
\eta_d=\frac{-q^2}{4m_d^2},
\label{abq}
\end{multline}
The cross section for elastic $ed$ scattering is written as
\begin{eqnarray}
\frac{d\sigma_{ed}}{d\Omega}=
\frac{\sigma_{Mott}}{1+\frac{2E}{m_d}sin^2\frac{\theta}{2}}
\left\{A(q^2)+B(q^2)\,tan^2\frac{\theta}{2}\right\},
\label{sigd}
\end{eqnarray}
The expression in the curly brackets of Eq.~(\ref{gdm}) evolves 
in the limit $Q^2\to$ 0 and it goes to the well known formula
for the deuteron magnetic moment
\begin{multline}
\mu_d(NN)=\mu_p+\mu_n-\frac{3}{2}(\mu_p+\mu_n-\frac{1}{2})P_D,\\
P_D=\int_{0}^{\infty}w^2(r)dr.
\label{dmagm}
\end{multline}

%%%%%%%%%%%%%%%%%%%%%%%%%%%%%%%%%%%%%%%%%%%%%%%%%%%%%%%%%%%%%%%%%
\section{The comparison with experimental data}
\label{sec:experiment}
%%%%%%%%%%%%%%%%%%%%%%%%%%%%%%%%%%%%%%%%%%%%%%%%%%%%%%%%%%%%%%%%%

When calculating the dibaryon- and quark contributions to both 
physical processes one begins from general formula (\ref{m1})
for the M1-amplitude and modifies its spin-dependent part. The
contribution of the contact dibaryon-induced interaction can
be found if to replacing the isoscalar nucleon spin-current
operator in Eq.~(\ref{m1})
\begin{eqnarray}
\frac{ieq}{2m_N}\,\lambda(\mu_n\!+\!\mu_p)j_0(qr/2)\,
\vec\epsilon^{\,(\lambda)*}\cdot
\frac{\vec\sigma_p+\vec\sigma_n}{2}
\label{m1spin}
\end{eqnarray}     
with the respective spin-dependent operator for 
the dibaryon contact term~(\ref{fin}). Consequently, in the
left bra-vector of the matrix element (\ref{m1}) one needs to use
the deuteron wavefunction $\Psi^{d}$  instead of $\Psi^{np}$. 
Moreover, one can ignore in this 
case the energy dependence of the non-local potential 
$V_{NqN}(r^{\prime},r;E)$ in Eq.~(\ref{sprb}) and substitute
$E=$ 0 there instead of the real energy $\varepsilon_T$ of thermal 
neutrons or the bound-state energy of deuteron $E_d$ since
the scale of the $\lambda(E)$ factor in Eqs.~(\ref{sprb}) - 
(\ref{lame}) is of much  larger and of the order $E_0\sim$ 1 GeV.
With these reasonable approximations one calculates first of
all the isoscalar current contribution to the deuteron magnetic
moment.

%%%%%%%%%%%%%%%%%%%%%%%%%%%%%%%%%%%%%%%%%%%%%%%%%%%%%%%%%%%%%%%%%%%
\subsection{The deuteron magnetic moment and the deuteron form factor}
%%%%%%%%%%%%%%%%%%%%%%%%%%%%%%%%%%%%%%%%%%%%%%%%%%%%%%%%%%%%%%%%%%%

In our model the deuteron magnetic form factor takes the form:
\begin{widetext}
\begin{multline}
G_M^d(q^2)=\sqrt{\frac{2}{3}}\frac{\sqrt{-q^2}}{2M_{\scriptscriptstyle N}}
\left\{cos^2\theta_{{\scriptscriptstyle N}q}G_{M(NN)}^d(q^2)+
cos^2\theta_{{\scriptscriptstyle N}q}\,
\mu_{{\scriptscriptstyle N}q{\scriptscriptstyle N}}
F_{{\scriptscriptstyle N}q{\scriptscriptstyle N}}(q^2)
+sin^2\theta_{{\scriptscriptstyle N}q}\,\mu_{s^6}F_{s^6}(q^2)\right.\\
\left.+2cos\,\theta_{{\scriptscriptstyle N}q}
sin\,\theta_{{\scriptscriptstyle N}q}\,\mu_{s^6-s^4p^2}F_{s^6-s^4p^2}(q^2)
\right\},
\label{mtotff}
\end{multline}
\end{widetext}
where the first term in the brackets represents the nucleonic current 
contribution while the second one corresponds to the 
isoscalar component of contact $NN\Leftrightarrow NN\gamma$ vertex (\ref{fin})
\begin{multline}
\mu_{{\scriptscriptstyle N}q{\scriptscriptstyle N}}
F_{{\scriptscriptstyle N}q{\scriptscriptstyle N}}(q^2)=
G^S_M(q^2)\,2Z\!\int_0^{\infty}\!\int_0^{\infty}drdr^{\prime}
u(r)u(r^{\prime})\\
\times\varphi_{2S}(r)\frac{\lambda(0)}{2M_N}
\frac{1}{q}j_1(qr^{\prime}/2)
\frac{d\varphi_{2S}(r^{\prime})}{dr^{\prime}}.
\label{munqn}
\end{multline}
Here, $F_{{\scriptscriptstyle N}q{\scriptscriptstyle N}}(0)=$ 1 by
definition and thus the value 
$\mu_{{\scriptscriptstyle N}q{\scriptscriptstyle N}}$ is equal to that
of right-hand side integral at $q=$ 0. The third and fourth terms 
in Eq.(\ref{mtotff}) represent the diagonal and non-diagonal contributions 
of the 6q-core of the dibaryon, i.e. the bare dibaryon contribution. 
As is evident from Eq.~(\ref{ndbff}) of Appendix~\ref{sec:appendc}
the last term in Eq.~(\ref{basic1})
vanishes at $q=$ 0 and thus does not contribute to the deuteron 
magnetic moment, while the account of the second term in 
Eq.~(\ref{basic1}) results only in a minor renormalization of the
deuteron magnetic moment (\ref{dmagm}). As a result, the dressed 
bag gives a real contribution to the deuteron magnetic moment only
due to contact $NN\Leftrightarrow NN\gamma$-vertex (\ref{fin}), and this 
contribution is equal to
\begin{eqnarray}
\Delta\mu^{DB}_d=cos^2\theta_{{\scriptscriptstyle N}q}\,
\mu_{{\scriptscriptstyle N}q{\scriptscriptstyle N}}.
\label{mudb}
\end{eqnarray}
In all the present calculations for the deuteron magnetic moment 
and the structure function $B(Q^2)$ the published parameters~(\ref{mtmod}) 
of the Moscow-Tuebingen $NN$-model have been employed. These 
parameters allow to fit the $NN$ phase shifts in the very large energy 
interval 0 - 1000 MeV. The mixing parameter 
$\theta_{{\scriptscriptstyle N}q}$ can also be calculated with the MT model 
amounting to
\begin{eqnarray}
sin\theta_{{\scriptscriptstyle N}q}=-0.13886.
\label{sin}
\end{eqnarray}
The first term in Eq.~(\ref{mtotff}) is calculated with Eq.~(\ref{dmagm}),
while the third and fourth terms are calculated via Eqs.~(\ref{dbff}) -
(\ref{ndbff}) of Appendix~\ref{sec:appendc}. The sum of these three terms to
the deuteron magnetic moment amounts to
\begin{multline}
\mu_d=cos^2\theta_{{\scriptscriptstyle N}q}\,\mu_d(NN)
+sin^2\theta_{{\scriptscriptstyle N}q}\,\mu_d(6q)\\=0.8489 n.m.,
\label{low}
\end{multline}
as shown in Table 1.
From the difference of this theoretical prediction to the respective
experimental value $\mu_d^{exp}=$ 0.8574 n.m. one can find an admissible 
value for the contact-term contribution Eq.~(\ref{mudb}), which in our 
case amounts $\Delta\mu^{DB}_{d}\sim$ 0.01 n.m. The second
term in Eq.~(\ref{mtotff}) is calculated with Eqs.~(\ref{munqn}) -
(\ref{mudb}) and we employ the fixed values (\ref{mtmod}) and (\ref{sin}) and
$Z=$~1 in Eq.~(\ref{fin}). Thus, in this calculation for $\Delta\mu^{DB}_{d}$
we do not use any free parameters and reach a value
\begin{eqnarray}
\Delta\mu^{DB}_{d}\,=\, 0.0159,
\label{mucont}
\end{eqnarray}
which is in a very reasonable agreement with the above limitation. The
resulting value for the deuteron magnetic moment $\mu^{theor}_d=$~0.8648~n.m.
overshoots the respective experimental value a little bit, but the remaining
disagreement $\Delta\mu=$~0.0074~n.m. has decreased strongly.

Now, in order to reproduce exactly the deuteron magnetic moment, we fix 
the value of $Z$ as follows: $Z=$ 0.7. We consider the accurate
experimental value for the deuteron magnetic moment to give a stringent test
for any new isoscalar current contribution.  With the fixed renormalization
constant $Z=$ 0.7 we calculate the deuteron magnetic form factor and the 
circular polarization $P_\gamma$. We follow this strategy in order to obtain a
parameter-free estimate for the latter observables.

\begin{figure}
\vspace{1.4cm}
\begin{center}
\epsfig{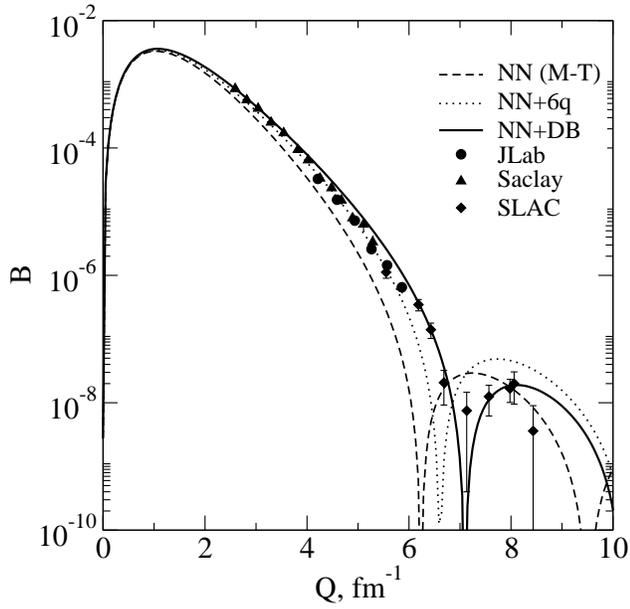}
\end{center}
\caption{The structure function $B(Q)$ of elastic e-d scattering.
The impulse approximation (IA) result for the Moscow-Tuebingen 
potential model~\cite{kuk1} is shown by the dashed line. 
The sum of IA and the diagonal ($s^6\to s^6$) and non-diagonal 
($s^6\to s^4p^2$) bare 6q-contributions is shown by the dotted line. 
The total contribution of the IA+ bare dibaryon + DB (contact term) 
contribution is represented by the solid line. The data are taken
from~\cite{sac} (Saclay), \cite{slac} (SLAC) and \cite{jlab} (JLab).}
\label{f9}
\end{figure}

Another restriction of the new isoscalar current is related to the description
of $B(Q^2)$ near its minimum at $Q^2\simeq$ 2~GeV$^2$/c$^2$ corresponding to
$Q\simeq$ 7~fm$^{-1}$. The position of the minimum depends crucially upon
non-additive two-body contributions.  In Fig.~\ref{f9} we display the results
of our calculation for $B(Q^2)$ based on Eq.~(\ref{mtotff}) and we compare
them to the experimental data~\cite{sac,slac,jlab}.  The dashed curve in
Fig.~\ref{f9} represents the single-nucleon current contribution which is
described by the term proportional to $G_{M(NN)}^d(q^2)$. The position of the
minimum for this single-nucleon term appears noticeably shifted toward lower
$Q^2$ values as compared to the experimental data~\cite{sac,slac,jlab}. Adding
the conventional quark contribution (dotted line in Fig.~\ref{f9}) reduces
this discrepancy due to the positive sign of the $s^6$-bag contribution which
is approximately compensated by the negative-sign interference term between
the nucleon and the bag contributions. It is evident however from this
consideration that one needs some positive contribution to reproduce the
correct position of the minimum.
 
%%%%%%%%%%%%%%%%%%%%%%%%%%%%%%%%% Fig.7 %%%%%%%%%%%%%%%%%%%%%%%%%%%%%%%%
%%%%%%%%%%%%%%%%%%%%%%%%%%%%%%%%%%%%%%%%%%%%%%%%%%%%%%%%%%%%%%%%%%%%%%%%
In the model developed here, the contact term which is tightly related to the
intermediate dibaryon production has just the necessary properties. 
Adding the contribution of the DB contact term (\ref{munqn}) in line
with Eq.~(\ref{mtotff}) results immediately in very good description 
for the deuteron magnetic form factor $B(Q^2)$ as shown by the solid line 
in Fig.~\ref{f9}. 
Thus, a rather minor renormalization of the $DB$ contact term by a factor of
0.7 makes it possible to describe quantitatively both the deuteron magnetic
moment~$\mu$ and the behavior of $B(Q^2)$ in the large momentum transfer
region $Q^2\lesssim$ 2,5~GeV$^2$/c$^2$. Finally, by fixing this minor
renormalization of the contact term the calculation the  circular
polarization $P_\gamma$ will be parameter-free.

%%%%%%%%%%%%%%%%%%%%%%%%%%%%%%%%%%%%%%%%%%%%%%%%%%%%%%%%%%%%%%%%%%%
\subsection{The circular polarization of photons 
in reaction $\vec n+p\to d+\vec\gamma$}
%%%%%%%%%%%%%%%%%%%%%%%%%%%%%%%%%%%%%%%%%%%%%%%%%%%%%%%%%%%%%%%%%%

The contribution of the dibaryon current to the isoscalar $M1$ transition
\mbox{$^3S_1(NN)\to^3S_1(d)$} is calculated now in the same way. When the
spin-dependent operator~(\ref{m1spin}) in the matrix element~(\ref{m1}) is
replaced by the contact term~(\ref{fin}) the $M1$-transition amplitude 
for the circularly polarized $\gamma$-quanta emission is obtained as
\begin{widetext}
\begin{multline}
\Delta M1^{(\lambda)}_{MM^{\prime}}=Z\int d^3r^{\prime}\int d^3r\,
\Psi^{d}_{M^{\prime}}(\vec r^{\,\prime})
\frac{e}{2M_N}i\left[
\frac{{\b\sigma}_p+{\b\sigma}_n}{2}\times{\bf q}
\right]\cdot{\b\epsilon}^{(\lambda)*}G^S_M(q^2)\\
\times\left\{\frac{1}{q}j_1(\frac{qr}{2})\frac{d\varphi_{2S}(r)}{dr}
\frac{\lambda(E_d)}{2M_N}\varphi_{2S}(r^{\prime})+
\varphi_{2S}(r)\frac{\lambda(\varepsilon_{\scriptscriptstyle T})}{2M_N}
\frac{1}{q}j_1(\frac{qr^{\prime}}{2})
\frac{d\varphi_{2S}(r^{\prime})}{dr^{\prime}}\right\}
\Psi^{np}_M(\vec r,\vec p_{n}),
\label{dltm}
\end{multline}
\end{widetext}
When calculating $P_{\gamma}$ this amplitude must be added to the
single-nucleon current terms~(\ref{mpg}) using the same 
renormalization constant $Z=$ 0.7.
 Similarly to the  single-nucleon current, the integral~(\ref{dltm}) is
 calculated 
straight-forward by replacement of $j_1(qr/2)/q\to r/6$ by
$q=E_d$. Also we  substitute $E=$~0 instead of $E=\varepsilon_T$
and $E=E_d$ in the $\lambda(E)$ function in Eq.~(\ref{lame}).
After this we get for the dibaryon induced current contribution
an expression analogous to Eq.~(\ref{m1s0}) with the respective
``reduced'' dibaryon matrix elements. The amplitude~(\ref{dltm})
found by this way together with the single-nucleon matrix elements
(\ref{mpg}) should be included to the final expression for 
$P_{\gamma}$
\begin{eqnarray}
P^{tot}_{\gamma}=2Re\left\{\frac{M1_0}{M1_1}+\frac{\Delta M1}{M1_1}
+\frac{E2_0}{M1_1}\right\},
\label{pgmmd}
\end{eqnarray}
where the dibaryon induced current contribution is
\begin{eqnarray}
P_{\gamma}(DB)=2Re\left\{
\frac{\Delta M1}{M1_1}\right\}.
\label{pgdb}
\end{eqnarray}
The results of the numerical calculations within our model are presented in
Table~\ref{tab:polar} together with a parallel calculation for $P_{\gamma}$
with the conventional RSC $NN$-potential model in its modern version
RSC93~\cite{e}. Evidently  the fully parameter free prediction of our
dibaryon model for $P_{\gamma}$ is in a first time in very good agreement 
with the respective
experimental result.

\begin{table*}
\caption{\label{tab:polar}
Circular polarization of $\gamma$ quanta 
in the $\vec n+p\to d+\vec\gamma$ reaction}
\begin{tabular}{|c|c|c|c|c|c|}
\hline
&&&&&\\[-1pt]
&$P_{\gamma}(M1)$&$P_{\gamma}(E2)$&$P_{\gamma}(NN)$&
$P_{\gamma}(DB)$&$P^{tot}_{\gamma}$\\[2pt]
Model&$\cdot10^{-3}$&$\cdot10^{-3}$&$\cdot10^{-3}$&$\cdot10^{-3}$&
$\cdot10^{-3}$\\[2pt]
\hline
&&&&&\\[-1pt]
Reid 93&-1.761&0.699&-1.062&0.&-1.062\\[2pt]
\hline 
Moscow-Tuebingen&-1.791&0.657&-1.134&-0.261&-1.395\\[2pt]
\hline
Experiment~\cite{lob}&&&&&-1.5$\pm$0.3\\[2pt]
\hline
\end{tabular}
\end{table*}

%%%%%%%%%%%%%%%%%%%%%%%%%%%%%%%%%%%%%%%%%%%%%%%%%%%%%%%%%%%%%%%%%%%%
\section{Short discussion and conclusion}
\label{sec:discussion}
%%%%%%%%%%%%%%%%%%%%%%%%%%%%%%%%%%%%%%%%%%%%%%%%%%%%%%%%%%%%%%%%%%%%%
In this paper we developed a model for the new electromagnetic current in the
deuteron and in the $NN$ system in general. The new currents are based on the
picture of short-range $NN$ interaction via an intermediate dibaryon
generation.  The dibaryon represents a new degree of freedom and according to
a general principle of quantum theory this must inevitably lead to the
respective new current(s).  By applying the general recipe of minimal
substitution to the Hamiltonian of the dibaryon model to derive the new
current one gets automatically two different contributions: diagonal and
transitional ones. The diagonal current is associated mainly with the quark
degrees of freedom, and thus is proportional to the (small) weight of the
dibaryon component in the deuteron.  While the transitional current leads to a
larger contribution to the deuteron electromagnetic properties, and likely also $NN$
electromagnetic observables, especially of isoscalar nature. We studied three
such electromagnetic  characteristics:
\begin{itemize}
\item the magnetic moment $\mu_d$ of the deuteron;
\item the magnetic form factor $B(Q^2)$ in the region of its diffraction 
minimum;
\item the circular polarization $P_{\gamma}$  of $\gamma$ quanta in
radiative capture of spin-polarized neutrons by hydrogen.
\end{itemize}
As for the prediction of the deuteron magnetic moment, the new isoscalar
dibaryon current just fills perfectly the small gap which was found earlier
($\Delta\mu\simeq$ 0.010 n.m.) between prediction of the dibaryon $NN$-force
model and experimental data (see Table~\ref{tab:deuteron}). With this tiny
correction the theoretical deuteron magnetic moment $\mu_d$ agrees excellently
with its respective experimental value.

In the present study we found that the minimal (gauge) substitution to the
dibaryon Hamiltonian gives a strong positive contribution to the $B(Q^2)$
behavior near the minimum region.  Moreover, the parameter-free calculation
of the $B(Q^2)$ in the new model gave already a very reasonable description
for the deuteron magnetic form factor $B(Q^2)$.  A minor reduction of the
dibaryon-$\gamma$ vertex by a factor 0.7 results in an excellent agreement
with the data both for $\mu_d$ and $B(Q^2)$.

After fixing all parameters of the new model, we calculated the magnitude of
the circular polarization of photons in $\vec n+p\to d+\gamma$ capture at
thermal energy. This fully parameter-free calculation gave a result which is
in a very close agreement with the existing experimental data~\cite{lob}.  It
is important to remind the reader that many attempts were undertaken in the
past; see e.g. the review of M. Rho~\cite{rho4} where one can find the
references to earlier works and a good discussion of all difficulties
encountered in theoretical predictions of $P_{\gamma}$. Thus, this
longstanding $P_{\gamma}$-puzzle seems now to be solved.

Here it is useful to discuss briefly the comparison between the present model
predictions and some other current models, both microscopic and
phenomenological ones. Very detailed six-quark microscopic calculations in
Ref.~\cite{buch} have revealed that the quark exchange currents 
cannot give
any quantitative agreement with deuteron data neither for the magnetic nor for
the charge form factors, $B(Q^2)$ and $A(Q^2)$, respectively. Moreover, when
calculating the quark-exchange current corrections to the magnetic and
quadrupole deuteron moments the authors~\cite{buch} have found some (although
small) underestimation for $\mu_d$ but strong {\it overestimation} for $Q_d$.
These disagreements with the respective experimental results have demonstrated
that the incorporation of a bare six-quark contribution only cannot fill the
gap between the impulse approximation (plus the traditional MEC) results and
the experiment, at least for the $M1$- and $E2$-isoscalar transitions. On the
other hand, the dressing procedure for six-quark bag has been shown in the
present work to lead inevitably to new short-range currents. These dibaryon
induced currents should replace the conventional two-body meson-exchange
currents at short $NN$-distances when two interacting nucleons are overlapping
strongly in which case their meson clouds will fuse into one common cloud of a
dibaryon.

The new dibaryon currents proposed and studied in this paper must contribute
also to many other electromagnetic  properties which could not be explained with the
conventional $NN$-models before, e.g. the $\gamma$-induced polarization of
nucleons at photo-disintegration of deuteron at low energies,
$d(\gamma,\vec{n})p$, and also the electro-disintegration of deuteron,
$d(e,e^{\prime}p)n$ at high momentum transfer~\cite{strok}. The particular
interest for the new isoscalar current rests in numerous studies of
($e,e^{\prime}pp$) and $(\gamma,pp)$ processes at intermediate energies. It is
worth to remember here that the theoretical interpretation of such processes,
measured experimentally at various kinematic conditions, failed to explain
the accurate experimental findings (see e.g. Ref.~\cite{groep}). Very likely
these processes include some contribution of two-body isoscalar currents as
well.

Simultaneously, any success in such consistent interpretation of the data will
support strongly the underlying dibaryon model for the short range $NN$
interaction.

\bigskip

\centerline{\bf Acknowledgments}
The authors appreciate very much the discussions with Profs. 
S.V. Gerasimov and V.E. Lyubovitskij. We are very
thankful to Drs.~V.N.~Pomerantsev, M.A. Shikhalev and M. Kaskulov
for very fruitful discussions and help.

\newpage

%\vspace{0.7cm}

%\centerline{\bf Appendix}

\appendix

\section{Avoiding double-counting}
\label{sec:appenda}
It is worth to add here some important comments about a large difference
between ``diagonal'' and ``non-diagonal'' (transition) currents in the $NN$
system. It is important that the similar consideration of the $t$-channel
two-body currents in the $NN$ system (with replacement quarks by nucleons in
Fig.~\ref{f6}) within the framework of a traditional $NN$ model does not lead
to new currents because the scalar exchange in the $t$-channel is already
included somehow into the $NN$ wavefunction $\Psi_{NN}$, and thus the
one-nucleon current matrix element
$\langle\Psi_{NN}|J^{\mu}_{N}|\Psi_{NN}\rangle$ also includes among the others
the diagrams shown in Fig.~\ref{f6}.

In contrast to this (one-channel) problem, we are dealing here with a two- (or
more) channel problem and our non-diagonal current operator $V_{Nq\gamma}$
given in Eq.~(\ref{vngq}) describes electromagnetic  induced transition between two
different channels, i.e. between the proper $NN$- and dressed bag components.
In such transitions there happens a strong rearrangement of the spin-isospin
structure of the total six-quark wavefunction, and thus this transition
between two components is associated with the real current of magnetic type,
which corresponds to the quark spin (isospin) flip.  It can be illustrated
clearly by considering a transition
$|s^4p^2[42]_X\rangle\rightarrow|s^6[6]_X\rangle$ even with the same value of
the total spin S (and isospin T) in the left (bra) and right (ket) vectors in
the current matrix element:
$\langle{s^6}[6]_X,ST|V_{Nq\gamma}|s^4p^2[42]_X,ST\rangle$.  Recall also that
the configuration $s^6$ in our model is fully absent in the initial and final
$NN$ states because of the orthogonality condition~(\ref{ortcnstr}) and the
projector appearing in the $NN$ channel.

Let us assume that the spin and isospin in the initial (and final)
$NN$-channel and in the intermediate dibaryon state have the values $S=$ 1 and
$T=$ 0, which are associated with the spin and isospin Young tableaux for the
six-quark system $[42]_S$ and $[3^2]_T$ respectively. As a result, we get the
following transition matrix element
\begin{multline}
\langle s^6[6]_X[42]_S[3^2]_T([f^{\prime}_{ST}]),[2^3]_C|\\
\times V_{Nq\gamma}\,
|s^4p^2[42]_X[42]_S[3^2]_T([f_{ST}]),[2^3]_C\rangle 
\label{stndtr}
\end{multline}
It is important to stress here that $[f^{\prime}_{ST}]\ne[f_{ST}]$, i.e. the
spin-isospin structure in bra- and ket states are different, because the Pauli
principle severely restricts the form of the spin-isospin tableau
$[f^{\prime}_{ST}]$ on the left hand side state reducing it to a single
allowable state $[42]_S\cdot[3^2]_T\to[f^{\prime}_{ST}]=[3^2]_{ST}$. In
contrast to it, on the right hand side state any spin-isospin Young tableau
$[42]_S\cdot[3^2]_T\to[f_{ST}]=
[51]_{ST}\!+\![41^2]_{ST\!}+\![3^2]_{ST}\!+\![321]_{ST}\!+\![2^21^2]_{ST}$ is
permissible (see e.g. Refs.~\cite{obu0}).
 
Therefore even the static magnetic moments for left and right functions must
be different. It is worth to remember here that just the spin-isospin Young
tableau $[f]_{ST}$ determines fully the nucleon magnetic moment $\mu_N$ at
$S=$ 1/2 and $T=$ 1/2, and the correct value $\mu_N$ both for proton and
neutron can be obtained only for the symmetric ST-state $[3]_{ST}$ satisfying
the Pauli exclusion principle.  Hence, in the transitions~(\ref{stndtr}) we
have real currents describing the spin-isospin flip in the $NN\to DB$
transition, and thus, the current operator $V_{Nq\gamma}$ makes a non-trivial
(two-body) contribution to the total current in the two-channel system
$NN+DB$.

\section{Derivation of the contact $NN\Leftrightarrow NN\gamma$ term}
\label{sec:appendb}

In deriving of Eq.~(\ref{fin}) we start from Eq.~(\ref{cnstr}) and 
substitute the vertex $Nq\gamma$ (\ref{vngq}) in one of two matrix
elements of the $s^4p^2\to s^6$ transition 
$<s^4p^2\{f\}|V_{Nq}|s^6[6]_{\scriptscriptstyle X}>$
in the r.h.s. of Eq.~(\ref{cnstr}). It gives
\begin{widetext}
\begin{multline}
V_{NqN\gamma}\simeq
\sum_{ff^{\prime}}\left[\langle \!NN|s^4p^2\{f\}\rangle\,\langle s^4p^2\{f\}|
V_{Nq\gamma}|s^6\rangle\, G_{DB}\,
\langle s^6|V_{Nq}|s^4p^2\{f^{\prime}\}\rangle\,\langle s^4p^2\{f^{\prime}\}|
NN\rangle\right.\\
\left.+\langle NN|s^4p^2\{f\}\langle\,\rangle s^4p^2\{f\}|
V_{Nq}|s^6\rangle\, G_{DB}\,
\langle s^6|V_{Nq\gamma}|s^4p^2\{f^{\prime}\}\langle\,\rangle s^4p^2
\{f^{\prime}\}|NN\rangle\right]
\label{nqg}
\end{multline}
\end{widetext}
(non-significant details are omitted here). The calculation of this operator
is performed with the fractional parentage coefficient technique (see
Refs.~\cite{obu1,obu0,obu2} for details) by factorizing the fixed pair $ij$ of
quarks with numbers $i=$~3 and $j=$~6. The space coordinates of this pair
depend on proton and neutron center-of-mass coordinates, ${\bf r}_p$ and ${\bf
  r}_n$ respectively, and on the relative motion Jacobi coordinates
${\b\rho}_{p1}$, ${\b\rho}_{p2}$, $\b\rho_{n1}$ and ${\b\rho}_{n2}$, as
defined in Eq.~(\ref{n123}) and below 
\begin{multline}
{\bf r}_3={\bf R}+{\bf r}_p-2{\b\rho}_{p2}/3,\quad 
{\bf r}_6={\bf R}+{\bf r}_n-2{\b\rho}_{n2}/3,\\
{\bf R}=\frac{1}{6}\sum_{i\!=\!1}^6{\bf r}_i.
\label{r36}
\end{multline}
In the c.m. frame ${\bf R}=$ 0 and ${\bf r}_p=-{\bf r}_n={\bf r}/2$. 
The conjugated  momenta ${\bf p}_3$ and ${\bf p}_6$ reads
\begin{multline}
{\bf p}_3={\bf P}+2({\bf p}_p-{\bf p}_n)/3-{\b\pi}_{p2},\\
{\bf p}_6={\bf P}-2({\bf p}_p-{\bf p}_n)/3-{\b\pi}_{n2}, 
\label{p36}
\end{multline}
where ${\b\pi}_{p2}$ and ${\b\pi}_{n2}$ are the relative momenta 
conjugated to ${\b\rho}_{p2}$ and ${\b\rho}_{n2}$ respectively,
${\bf P}=\sum_{i\!=\!1}^6{\bf p}_i$.

The calculation of the Fourier transform
$\int\frac{d^3p_1}{(2\pi)^3}\dots\int\frac{d^3p_6}{(2\pi)^3} exp(i{\bf
  p}_1\cdot {\bf r}_1+\dots+i{\bf p}_6\cdot {\bf r}_6)$ for the matrix
elements of operators $V_{Nq\gamma}$ and $V_{Nq}$ on the r.h.s. of
Eq.~(\ref{nqg}) and substitution of the integral $\int exp\{-i({\bf p}_3+{\bf
  p}_6-{\bf p}^{\prime}_3- {\bf p}^{\prime}_6-{\bf q})\cdot{\bf x}\} d{\bf x}$
for the $\delta$-function $(2\pi)^3\delta^3({\bf p}_3+{\bf p}_6-{\bf
  p}^{\prime}_3- {\bf p}^{\prime}_6-{\bf q})$ lead, after the simple but
tedious mathematics, to the following expression for the $r$-dependent part of
the $NN\to DB+\gamma$ vertex in Eq.~(\ref{nqg})
\begin{widetext}
\begin{equation}
\tilde V_{Nq\gamma}(q,r)=\frac{e}{2M_{\scriptscriptstyle N}}
F_{\scriptscriptstyle N}({\bf q}^2)\,g_s^2\,\langle v\rangle\, 
\left\{e^{-i{\bf q}\cdot{\bf r}/2}\mu_p
\frac{1}{M_{\scriptscriptstyle N}}{\b\nabla}_r\cdot
[{\b\epsilon}^{(\lambda)}\times{\b\sigma}_p]
-e^{i{\bf q}\cdot{\bf r}/2}\mu_n
\frac{1}{M_{\scriptscriptstyle N}}{\b\nabla}_r\cdot
[{\b\epsilon}^{(\lambda)}\times{\b\sigma}_n]\right\}\varphi_{2S}(r).
\label{cntmag}
\end{equation}
\end{widetext}
The terms with factors
$\frac{\hat{\bf q}\cdot{\bf k}_i}{m_q}[{\b\sigma}_i\times\hat{\bf q}]$
on the r.h.s. of Eq.~(\ref{vngq}) are canceled after the summation 
in the Eq.~(\ref{nqg}).
Now the expression (\ref{cntmag}) should replace the function 
$\varphi(r)\varphi_{2S}(r)$ in the separable 
potential~(\ref{sprb}) in line with
the interpretation of the vertex matrix element in Eq.~(\ref{2s}).

In Eq.~(\ref{cntmag}) we use the standard formulas of CQM for the 
nucleon form factor
\begin{eqnarray}
F_{\scriptscriptstyle N}({\bf q}^2)=F_q(q^2)
\int|\psi_N(\rho_1,\rho_2))|^2
e^{i2{\b\rho}_2\cdot {\bf q}/3}d^3\rho_1d^3\rho_2
\label{nff}
\end{eqnarray}
and for the magnetic momentum of nucleon 
\begin{multline}
\langle N(3q),[21]_{\scriptscriptstyle S}[21]_{\scriptscriptstyle T}:
[3]_{\scriptscriptstyle ST}|\\
\times\sum_{i\!=\!1}^3\frac{\hat e_i{\b\sigma}_i}{2m_q}
|N(3q),[21]_{\scriptscriptstyle S}
[21]_{\scriptscriptstyle T}:[3]_{\scriptscriptstyle ST}\rangle\,
=\,\frac{\mu_{\scriptscriptstyle N}{\b\sigma}_{\scriptscriptstyle N}}
{2M_{\scriptscriptstyle N}},
\label{nmag}
\end{multline}
$N\!=\!n,p$, with $\mu_p=$~3, $\mu_n=$~-2. The gradients in Eq.~(\ref{cntmag})
originate from the proton and neutron momenta ${\bf k}_p$ and ${\bf k}_n$
which appear in the momentum representation of the vertex (\ref{cntmag}) as a
result of substitution of Eqs. (\ref{p36}) for the quark momenta ${\bf p}_3$
and ${\bf p}_6$.  In the coordinate representation, these momenta transform
into gradients $-i{\b\nabla}_{r_p}=-i{\b\nabla}_{r/2}$ and
$-i{\b\nabla}_{r_n}=i{\b\nabla}_{r/2}$ respectively. The nucleon mass
$M_{\scriptscriptstyle N}$ in Eq.~(\ref{nmag}) is a result of the substitution
of $M_{\scriptscriptstyle N}$ for the value 3$m_q$ which appears in the r.h.s.
of Eq.~(\ref{nqg}) resulting from CQM algebra.

Making use of the expressions
\begin{multline}
{\b\nabla}_{r}\varphi_{2S}(r)=\hat{\bf r}\frac{d}{dr}\varphi_{2S}(r),
\qquad e^{\pm i{\bf q}\cdot{\bf r}/2}=\\
j_0(qr/2)\pm i\sqrt{4\pi} j_1(qr/2)
\sum_{m}Y_{1m}(\hat q)Y_{1m}^*(\hat r)+\dots,
\label{grad}
\end{multline}
and adding the contributions of two graphs ($a$) and ($b$) depicted
in Fig.~\ref{f2} we obtain finally Eq.~(\ref{fin}) for the $NqN\gamma$ 
(contact) term.

\section{Nucleon and dibaryon electromagnetic matrix elements}
\label{sec:appendc}

The general formula for the nucleon electromagnetic current
\begin{multline}
eJ_{\scriptscriptstyle N}^{\mu}=e_{\scriptscriptstyle N}\gamma^{\mu}
F_1^{\scriptscriptstyle N}(q^2)+
i\frac{e\mu_{\scriptscriptstyle N}-e_{\scriptscriptstyle N}}
{2M_{\scriptscriptstyle N}}\sigma^{\mu\nu}q^{\nu}     
F_2^{\scriptscriptstyle N}(q^2),\\
e_{\scriptscriptstyle N}=e(\frac{1}{2}+\frac{\tau_z}{2})
\label{ejn}
\end{multline}
can easily be derived after averaging the quark current (\ref{jq})
over the nucleon wavefunction:
\begin{multline}
eJ_{{\scriptscriptstyle N}QM}^{\mu}=
\langle N(123)|\sum_{i\!=\!1}^3j^{\mu}_i(q)|N(123)\rangle\\
=\left[e_{\scriptscriptstyle N}\gamma^{\mu}+
i\frac{e\mu_{\scriptscriptstyle NQM}-e_{\scriptscriptstyle N}}
{2M_{\scriptscriptstyle N}}\sigma^{\mu\nu}q^{\nu}\right]
F_{QM}(q^2),
\label{ejqm}
\end{multline}
in which the quark-model predictions
\begin{multline}
\mu_{p{\scriptscriptstyle QM}}=3,\quad \mu_{n{\scriptscriptstyle QM}}=-2,\\ 
F_{QM}(Q^2)=F_q(Q^2)e^{-Q^2b^2/6}    
\label{fqm}
\end{multline}
are not distinguished seriously from their experimentally measured
counterparts used in Eq.~(\ref{ejn}):
\begin{multline}
\mu_{p}=2.79,\quad \mu_{n}=-1.91,\\
F_2(Q^2)\approx F_{dip}(Q^2)=\left(\frac{1}{1+\frac{Q^2}{\Lambda_N^2}}    
\right)^2.
\label{nfqm}
\end{multline}
\medskip
%%%%%%%%%%%%%%%%%%%%%%%%%%%%% Fig. 8  %%%%%%%%%%%%%%%%%%%%%%%%%%%%%%%%%%
\begin{figure}
\vspace{0.5cm}
\begin{center}
\epsfig{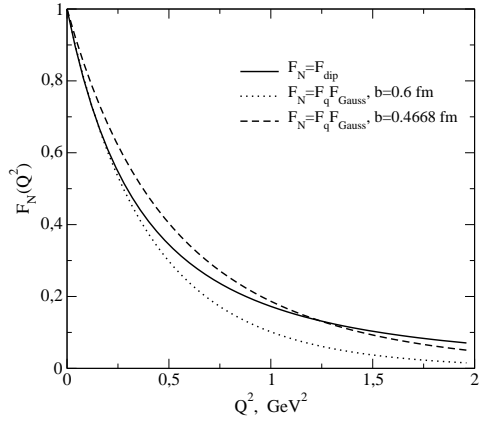}
\end{center}
\caption{Comparison of the simple quark model description for the
nucleon magnetic form factor (with different scale values $b$) with 
those parametrized through the standard dipole fit.}
\label{f8}
\end{figure}
%%%%%%%%%%%%%%%%%%%%%%%%%%%%%%%%%%%%%%%%%%%%%%%%%%%%%%%%%%%%%%%%%%%%%%%
It is important to stress here that the quark-model magnetic form factor,
as can be seen in Fig.~\ref{f8}, resembles quite precisely the respective
nucleon form factor found with the standard dipole fit, in the region
$Q^2\simeq$ 1 - 2 GeV$^2$/c$^2$, i.e. near the region of minimum for
the deuteron magnetic form factor $B(Q^2)$ --- see the respective 
curve in Fig.~\ref{f8} for the choice of
$b=\sqrt{\frac{3}{2}}r_0\simeq$ 0.47 fm. 

Thus, in the nucleon sector, we replace the quark-model current
$J_{{\scriptscriptstyle N}QM}^{\mu}$ with the standard representation
of the nucleon current $J_{{\scriptscriptstyle N}}^{\mu}$ given by 
Eqs.~(\ref{ejn}) and (\ref{nfqm}), and the non-additive
two-body current (which gives only a small correction to the
single-nucleon current $J_{{\scriptscriptstyle N}}^{\mu}$) only is 
calculated on the basis of the constituent quark model (CQM).

In particular, the CQM technique is used for calculation of the last 
two terms in Eq.~(\ref{basic1}). These terms are contributions of the 
graphs shown in Figs.~\ref{f5}(f) and (e) respectively.
\begin{itemize}
\item[($i$)] For the graph in Fig.~\ref{f5}(f), the 
diagonal matrix element 
$\langle DB|\sum_{i\!=\!1}^6 j^{\mu}_i\varepsilon_{\mu}^{(\lambda)}|DB\rangle$ 
is reduced here to 
the matrix element with the $s^6$ bag-like wavefunctions. This term
is found by the same technique as for the nucleon matrix element
calculation in Eqs.~(\ref{ejqm}) and (\ref{fqm}). As a result,
for the transverse current component ($\lambda=\pm$ 1, 
$\varepsilon_{\mu}^{(\lambda)}=\{0,{\b\epsilon}^{(\lambda)}\}$)
one gets
\begin{multline}
\langle DB|\sum_{i\!=\!1}^6j^{\mu}_i\varepsilon_{\mu}^{(\lambda)}|DB\rangle=\\
\langle s^6,S\!=\!1,\,T\!=\!0\,|\sum_{i\!=\!1}^6j^{\mu}_i
\varepsilon_{\mu}^{(\lambda)}\,|\,s^6,S\!=\!1,\,T\!=\!0\rangle\\
=\,-\,\frac{({\b\sigma}_p\!+\!{\b\sigma}_n)\cdot{\b\epsilon}^{(\lambda)}}{2}
\mu_{\scriptscriptstyle N}^sF_{s^6}(Q^2), 
\label{dbff}
\end{multline}
with $F_{s^6}(Q^2)=F_q(Q^2)e^{-5Q^2b^2/24},\quad
\mu_{\scriptscriptstyle N}^s=\mu_p+\mu_n$.
\item[($ii$)] With the graph in Fig.~\ref{f5}(e), the transition matrix 
element $NN\to DB$ can be found similarly to Eq.~(\ref{cnstr}), 
i.e. by using the expansion over the six-quark shell-model states with 
restriction by the most important low-lying states 
\end{itemize}
\begin{widetext}
\begin{multline}
\langle d(NN)|\sum_{i\!=\!1}^6j^{\mu}_i
\varepsilon_{\mu}^{(\lambda)}|DB\rangle\,=
\sum_f\int u(r)\langle N(123)|\langle N(456)|\,s^4p^2\{f\}\rangle dr\,
\langle s^4p^2\{f\}\,|\sum_{i\!=\!1}^6j^{\mu}_i\varepsilon_{\mu}^{(\lambda)}|
\,s^6\rangle\\
=\,-\,\frac{({\b\sigma}_p\!+\!{\b\sigma}_n)
\cdot{\b\epsilon}^{(\lambda)}}{2}\langle u|2S(NN)\rangle \sum_fC_f
F_{s^6-s^4p^2}(Q^2).
\label{ndbff}
\end{multline}
\end{widetext}
Here $F_{s^6-s^4p^2}(Q^2)=\frac{5Q^2b^2}{24}e^{-5Q^2b^2/24}$ and
$<u|2S(NN)>=\int_0^{\infty}u(r)\varphi_{2S}(r)dr$.
%\end{widetext}

\begin{widetext}
\section{Deuteron $M$1- and $E$2- transition amplitudes}
\label{sec:appendd}

The M1- and E2-amplitudes which are used for the cross section calculation
in Eqs.~(\ref{stot}) and (\ref{siglam}) take the following form (we 
present here them separately for the spin (s)- and convection (c) current 
components):
\begin{eqnarray}
M1^{(\lambda)I\!=\!1}_{\scriptscriptstyle MM^{\prime}}(s)=
\delta_{\scriptscriptstyle M^{\prime},-\lambda}
\delta_{\scriptscriptstyle M,0}\,(\mu_n-\mu_p)\,
\lambda\left(\frac{-ieq}{2m_N}\right)
I_0(u,{ }^1S_0),
\label{m1s1}
\end{eqnarray}
%--------------------------
\begin{multline}
M1^{(\lambda)I\!=\!0}_{\scriptscriptstyle MM^{\prime}}(s)=
\delta_{\scriptscriptstyle M^{\prime},M-\lambda}\,(\mu_n+\mu_p)\,
[\lambda\sqrt{2}(1M1-\!\lambda)|1(M\!-\!\lambda)]
\left(\frac{-ieq}{2m_N}\right)\\
\times\left\{I_0(u,{ }^3S_1)-\frac{1}{2}I_0(w,{ }^3D_1)+\frac{1}{\sqrt{2}}
I_2(uw,{{ }^3S_1}{ }^3D_1)+\frac{1}{2}I_2(w,{ }^3D_1)\right\},
\label{m1s0}
\end{multline}
%-------------------------
\begin{eqnarray}
M1^{(\lambda)I\!=\!0}_{\scriptscriptstyle MM^{\prime}}(c)&=&
\delta_{\scriptscriptstyle M^{\prime},M-\lambda}\,
[\lambda\sqrt{2}(1M1-\!\lambda)|1(M\!-\!\lambda)]
\left(\frac{-ieq}{2m_N}\right)
\frac{3}{8}\left\{I_0(w,{ }^3D_1)+I_2(w,{ }^3D_1)\right\},
\label{m1c}
\end{eqnarray}
%-------------------------
\begin{eqnarray}
E2^{(\lambda)}_{\scriptscriptstyle MM^{\prime}}(s)&=&
\delta_{\scriptscriptstyle M^{\prime},M-\lambda}(\mu_n+\mu_p)
\left[-\frac{\sqrt{10}}{\sqrt{3}}(1M2-\!\lambda)|1(M\!-\!\lambda)\right]
\left(\frac{-ieq}{2m_N}\right)
\frac{3}{\sqrt{2}}I_2(uw,{{ }^3S_1}{ }^3D_1),
\label{e2s}
\end{eqnarray}
%------------------------
\begin{eqnarray}
E2^{(\lambda)}_{\scriptscriptstyle MM^{\prime}}(c)&=&
\delta_{\scriptscriptstyle M^{\prime},M-\lambda}\,
[\lambda\sqrt{2}(1M1-\!\lambda)|1(M\!-\!\lambda)]
\left(\frac{-ieq}{2m_N}\right)
\frac{3}{8}\left\{I_0(w,{ }^3D_1)+I_2(w,{ }^3D_1)\right\},
\label{e2c}
\end{eqnarray}
where $I_0$ and $I_2$ are the following overlap integrals
\begin{eqnarray}
&I_0(f,Z)=\int_0^{\infty}f(r)Z(r,p_n)j_0(qr/2)dr,\qquad
I_2(f,Z)=\int_0^{\infty}f(r)Z(r,p_n)j_2(qr/2)dr,&\nonumber\\
&I_2(uw,{{ }^3S_1}{ }^3D_1)=\frac{1}{2}\int_0^{\infty}
\left[u(r){ }^3D_1(r,p_n)+w(r){ }^3S_1(r,p_n)\right]j_2(qr/2)dr.&
\label{i0i2}
\end{eqnarray}
Here $Z(r,p_n)$ can be any of the scattering wavefunctions in $^1S_0$,
$^3S_1$ or $^3D_1$ channels.

For the sake of convenience, the factors which are equal to unity 
on modulo are separated out in square brackets:
$$
\lambda\sqrt{2}(1M1-\!\lambda)|1(M\!-\!\lambda)=1,\qquad
-\frac{\sqrt{10}}{\sqrt{3}}(1M2-\!\lambda)|1(M\!-\!\lambda)=(-1)^M.
$$
 When summing over the $M$ and $\lambda$ induces, these terms play a role
similar to the Kroneker delta.
\end{widetext}

%%%%%%%%%%%%%%%%%%%%%%%%%%%%%%%%%%%%%%%%%%%%%%%%%%%%%%%%%%%%%%%%%%%%
\end{document}